\documentclass[showpacs,prd,twocolumn]{revtex4}
\usepackage{pifont}
\usepackage{bbding}
\usepackage{mathrsfs}
\usepackage{longtable,lscape} \usepackage{multirow}
\usepackage{txfonts,bm}
\usepackage{amssymb}
\usepackage{indentfirst}
\usepackage{graphicx,booktabs}
\usepackage{color}
\usepackage{hyperref}

\def\f{\frac}

\def\L{\Lambda}

\def\hright{{\smallsetminus\!\!\!\!\!\! \surd}}

\begin{document}

\title{Coupled-channel analysis of the possible $D^{(*)}D^{(*)}$,
$\bar{B}^{(*)}\bar{B}^{(*)}$ and $D^{(*)}\bar{B}^{(*)}$ molecular
states}
\author{Ning Li$^{1,2}$}\email{leening@pku.edu.cn}
\author{Zhi-Feng Sun$^{3,4}$}\email{sunzhif09@lzu.edu.cn}
\author{Xiang Liu$^{3,4}$}\email{xiangliu@lzu.edu.cn}
\author{Shi-Lin Zhu$^{1,5,6}$}\email{zhusl@pku.edu.cn}
\affiliation{$^1$Department of Physics
and State Key Laboratory of Nuclear Physics and Technology, Peking University, Beijing 100871, China\\
$^2$Institut f\"{u}r Kernphysik and J\"ulich Center for Hadron Physics, Forschungszentrum J\"{u}lich, D-52425 J\"{u}lich, Germany\\
$^3$School of Physical Science and Technology, Lanzhou University, Lanzhou 730000,  China\\
$^4$Research Center for Hadron and CSR Physics, Lanzhou University
and Institute of Modern Physics of CAS, Lanzhou 730000, China\\
$^5$Center of High Energy Physics, Peking University, Beijing
100871, China\\
$^6$Collaborative Innovation Center of Quantum Matter, Beijing 100871, China}

\date{\today}

\begin{abstract}

We perform a coupled-channel study of the possible deuteron-like
molecules with two heavy flavor quarks, including the systems of
$D^{(*)}D^{(*)}$ with double charm, $\bar{B}^{(*)}\bar{B}^{(*)}$
with double bottom and $D^{(*)}\bar{B}^{(*)}$ with both charm and
bottom, within the one-boson-exchange potential model. In our study, we take
into account the S-D mixing which plays an important role in the
formation of the loosely bound deuteron, and particularly, the
coupled-channel effect in the flavor space. According to our
results, the state $D^{(*)}D^{(*)}[I(J^P)=0(1^+)]$ with double charm, the states
$\bar{B}^{(*)}\bar{B}^{(*)}[I(J^P)=0(1^+),1(1^+)]$, $(\bar{B}^{(*)}\bar{B}^{(*)})_s[J^P=1^+,2^+]$ and $(\bar{B}^{(*)}\bar{B}^{(*)})_{ss}[J^P=1^+,2^+]$ with double
bottom, and the states $D^{(*)}\bar{B}^{(*)}[I(J^P)=0(1^+),0(2^+)]$ with both charm and bottom might be good molecule candidates. However, the states
$D^{(*)}D^{(*)}[I(J^P)=0(2^+),1(0^+),1(1^+),1(2^+)]$, $(D^{(*)}D^{(*)})_s[J^P=0^+,2^+]$ and $(D^{(*)}D^{(*)})_{ss}[J^P=0^+,1^+,2^+]$ with double charm and the state
$D^{(*)}\bar{B}^{(*)}[I(J^P)=1(1^+)]$ with both charm and bottom are not supported to be molecules.

\end{abstract}
\pacs{14.40.Rt, 14.40.Lb, 12.39.Hg, 12.39.Pn}

\maketitle

\section{Introduction} \label{Introduction}

More and more experimental observations have stimulated the
extensive discussions of exotic states. The molecular state
explanation to the reported charmonium-like states $X,Y,Z$ becomes
popular due to the fact that many charmonium-like states near the
threshold of charmed meson pair, {\it i.e.},
\begin{eqnarray*}
X(3872)&\sim& m_{D\bar{D}^*}, \quad
Y(3930)\sim m_{D^*\bar{D}^*}\\
Y(4140)&\sim& m_{D_s^*\bar{D}_s^*},\quad Y(4274)\sim
m_{D_{s0}(2317)\bar{D}_s}.
\end{eqnarray*}
And, in the past decade there is abundant literature with the
study of the heavy flavor molecular
states~\cite{Tornqvist:2003na,Close2004,Swanson2004,Tornqvist2004,
Liu:2005ay,Zhu2005,AlFiky:2005jd,Liu2008c,Liu2008,Liu2008a,Thomas:2008ja,
Lee:2008tz,Close2009,Ding2009a,Ding2009b,Liu2009c,Liu2009b,Liu2009a,
Lee:2008gn,Ortega2010,Ohkoda:2011vj,Ohkoda:2012uy,Aceti:2012qd}.

The concept of molecular state with hidden charm was first
proposed by Voloshin and Okun thirty years ago and they studied
the interaction between the charmed and anti-charmed
mesons~\cite{Voloshin1976}. Later, De Rujula, Georgi and Glashow
suggested that the observed $\psi(4040)$ is a $D^*\bar{D}^*$
molecule~\cite{DeRujula1977}. By the quark-pion interaction model,
T\"ornqvist investigated the possible deuteron-like two meson
bound states with $B\bar{B}^*$ or $B^*\bar{B}^*$
component~\cite{Tornqvist1994,Tornqvist1994a}. At present,
carrying out the phenomenological study of the heavy flavor
molecular state is still a hot research topic of hadron physics.

Usually, the hadron configurations mainly include
\begin{eqnarray*}
\mathrm{Hadron}\left\{
  \begin{array}{cccccc}
    \mathrm{Meson}: & q\bar{q},\quad Q\bar{q},\quad Q\bar{Q}\\
    \mathrm{Baryon}: & qqq,\quad Qqq,\quad QQq,\quad...\\
    \mathrm{Exotic\,state}:&\left\{
  \begin{array}{cccccc}
  \mathrm{Molecular\,state}\\
   \mathrm{Hybrid}\\
    \mathrm{Glueball}\\
    ...
  \end{array}
\right.
  &
  \end{array}
\right. ,
\end{eqnarray*}
where $q$ and $Q$ denote the light ($u,d,s$) and heavy ($c,b$)
quarks, respectively. Among the conventional baryon states, the
baryons with double charm or double bottom is of the $QQq$
configuration. The SELEX Collaboration reported the first
observation of a doubly charmed baryon $\Xi_{cc}^+$ in its charged
decay mode $\Xi_{cc}^{+}\to\Lambda_c^+K^-\pi^+$
\cite{Mattson:2002vu} and confirmed it in the decay mode
$\Xi_{cc}^{+}\to pD^+K^-$~\cite{Ocherashvili:2004hi}. However,
later the BABAR Collaboration searched for $\Xi^{+}_{cc}$ in the
final states $\Lambda_c^+K^{-}\pi^+$ and $\Xi_c^0\pi^{+}$, and
$\Xi_{cc}^{++}$ in the final states
$\Lambda_c^{+}K^{-}\pi^{+}\pi^{-}$ and $\Xi_c^{0}\pi^+\pi^+$, and
found no evidence for the production of the doubly charmed
baryons~\cite{Aubert:2006qw}. The Belle Collaboration reported no
evidence for the doubly charmed baryons in the final state
$\Lambda_c^{+}K^{-}\pi^{+}$, either~~\cite{Chistov:2006zj}.
Although these doubly charmed baryons were not confirmed by BABAR
and BELLE, it is still an interesting research topic to search for such doubly charmed baryons experimentally.

Besides the doubly heavy flavor baryons, it is also very
interesting to study other systems with two heavy flavor quarks.
The heavy flavor molecular state with two charm quarks provides
another approach to investigate the hadron states with double
charm. For this kind of hadron, its typical configuration is
$[c\bar{q}][c\bar{q}]$. To answer whether there exist such heavy
flavor molecular states with double charm or not, in this paper we
apply the one-boson-exchange (OBE) model to perform a dynamic
calculation of their mass spectroscopy. This study is not only a
natural extension of the previous work of the heavy flavor
molecular state with hidden charm, but also provides new insight
into exploring the hadron states with double charm. Besides the
hadron states with double charm, we also investigate the hadron
states with double bottom and the hadron states with both charm
and bottom.

This paper is organized as follows. After the introduction, we
present the derivation of the effective potential in
Section~\ref{formalism}. We summarize our numerical results and
perform some analysis in Section~\ref{Results} and  draw some
conclusions in Section~\ref{conclusion}. We also give some useful
formulas in the Appendix.

\section{Formalism}\label{formalism}

\subsection{The lagrangians and the coupling constants}

In the present paper, we investigate the possible molecules of
($D^{(*)}D^{(*)}$) with double charm,
($\bar{B}^{(*)}\bar{B}^{(*)}$) with double bottom and
($D^{(*)}\bar{B}^{(*)}$) with both charm and bottom. In our study,
we take into account the S-D mixing which plays an important role
in the formation of the loosely bound deuteron and, particularly,
the coupled-channel effects in the flavor space. We study the
systems with total angular momentum $J\leq2$. We list the channels
for different systems in
Tables~\ref{channel-DDBB}-\ref{channel-DB}.

\renewcommand{\arraystretch}{1.5}
\begin{table*}[htp]
\centering \caption{The different channels for the
$D^{(*)}D^{(*)}$ systems, and similarly, for the
$\bar{B}^{(*)}\bar{B}^{(*)}$. $S$ is the strangeness while $I$
stands for the isospin of the systems. ``$***$" means the
corresponding state dose not exist due to the symmetry. For
simplicity, we adopt the shorthand notations,
$[DD^*]_{-}\equiv{1\over\sqrt{2}}\left(DD^*-D^*D\right)$ and
$[DD^*]_{+}\equiv{1\over\sqrt{2}}\left(DD^*+D^*D\right)$.}
\label{channel-DDBB}
\begin{tabular*}{18cm}{@{\extracolsep{\fill}}ccccccccc}
\toprule[1pt]
        ~        &       ~     &   \multicolumn{7}{c}{Channels}
 \\
       $S$       &  $I(J^P)$   &     $1$     &     $2$     &
       $3$       &      $4$        &   $5$    &     $6$     &     $7$    \\
\hline
\multirow{6}*{0} &  $0(0^+)$   & \multicolumn{7}{c}{$***$}    \\
        ~        &  $0(1^+)$    &$[DD^*]_{-}(^3S_1)$&$[DD^*]_{-}(^3D_1)$ &
$D^*D^*(^3S_1)$  &$D^*D^*(^3D_1)$  &    -     &      -      &     -      \\
       ~         &  $0(2^+)$   &$[DD^*]_{-}(^3D_2)$&$D^*D^*(^3D_2)$     &
       -         &       -         &    -     &      -      &     -      \\ \cline{2-9}
      ~          &  $1(0^+)$   &   $DD(^1S_0)$     &$D^*D^*(^1S_0)$ &
$D^*D^*(^5D_0)$  &       -         &   -      &      -      &    -      \\
      ~          &  $1(1^+)$   &$[DD^*]_{+}(^3S_1)$&$[DD^*]_{+}(^3D_1)$ &
$D^*D^*(^5D_1)$  &       -         &   -      &      -      &    -      \\
      ~          &  $1(2^+)$   &$D^*D^*(^5S_2)$
&$D^*D^*(^1D_2)$  &$D^*D^*(^5D_2)$ &   -      &      -      &    -   & -   \\
\cline{1-9}
\multirow{3}*{$1$}& ${1\over 2}(0^+)$
&$DD_s(^1S_0)$      &$D^*D_s^*(^1S_0)$    &
$D^*D_s^*(^5D_0)$&       -         &    -     &      -      &    -      \\
      ~          &  ${1\over 2}(1^+)$   &$DD_s^*(^3S_1)$    &$DD_s^*(^3D_1)$ &$D^*D_s(^3S_1)$  &$D^*D_s(^3D_1)$  &$D^*D^*_s(^3S_1)$&$D^*D^*_s(^3D_1)$&$D^*D^*_s(^5D_1)$ \\
      ~          &  ${1\over 2}(2^+)$   &$D^*D_s^*(^5S_2)$
&$D^*D_s^*(^1D_2)$&$D^*D_s^*(^5D_2)$&  -     &      -      &     -   &  -  \\
\cline{1-9}
\multirow{3}*{$2$}& $0(0^+)$
&$D_sD_s(^1S_0)$    &$D_s^*D_s^*(^1S_0)$&$D_s^*D_s^*(^5D_0)$&
-                &  -    &     -   &   -     \\
       ~         &  $0(1^+)$   &$[D_sD_s^*]_{+}(^3S_1)$
&$[D_sD_s^*]_{+}(^3D_1)$&$D_s^*D_s^*(^5D_1)$&     -
&     -          &        -        &    -    \\
       ~         &  $0(2^+)$   &$D_s^*D_s^*(^5S_2)$
&$D_s^*D_s^*(^1D_2)$&$D_s^*D_s^*(^5D_2)$&   - &   -     &   -     &   -   \\
\bottomrule[1.0pt]
\end{tabular*}
\end{table*}

\renewcommand{\arraystretch}{1.5}
\begin{table*}[htp]
\centering \caption{The different channels  for the
$D^{(*)}\bar{B}^{(*)}$ systems. ``$S$" is the strangeness of the
corresponding system while ``$I$" stands for the isospin of the
state.}\label{channel-DB}
\begin{tabular*}{18cm}{@{\extracolsep{\fill}}cccccccccccc}
\toprule[1.0pt]
          ~      &        ~  &   \multicolumn{7}{c}{Channels}  \\
       $S$       & $I(J^P)$  &     $1$    &    $2$    & $3$ &
$4$              &      $5$        &     $6$   &    $7$    &
$8$              &      $9$        &$10$   \\
\hline \multirow{6}*{0} &$0(0^+)$    &
$D\bar{B}(^1S_0)$ &$D^*\bar{B}^*(^1S_0)$&
$D^*\bar{B}^*(^5D_0)$ &     -      &    -     &    -       &   -       &  -  &
       -         &      - \\
        ~        &  $0(1^+)$    &$D\bar{B}^*(^3S_1)$
&$D\bar{B}^*(^3D_1)$  &$D^*\bar{B}(^3S_1)$    &$D^*\bar{B}(^3D_1)$
&$D^*\bar{B}^*(^3S_1)$&$D^*\bar{B}^*(^3D_1)$  &$D^*\bar{B}^*(^5D_1)$ &  -  &
-                &     -  \\
       ~         &  $0(2^+)$
&$D^*\bar{B}^*(^5S_2)$&$D^*\bar{B}^*(^1D_2)$&$D^*\bar{B}^*(^5D_2)$   &  - &
    -            &     -           &     -    &   -      &     -     &  -  \\
\cline{2-12}
                 &  $1(0^+)$    & $D\bar{B}(^1S_0)$ &$D^*\bar{B}^*(^1S_0)$&
$D^*\bar{B}^*(^5D_0)$&     -       &    -     &   -     &    -     &   -  &
-                &     -     \\
        ~        &  $1(1^+)$    &$D\bar{B}^*(^3S_1)$&$D\bar{B}^*(^3D_1)$ &$D^*\bar{B}(^3S_1)$
&$D^*\bar{B}(^3D_1)$&$D^*\bar{B}^*(^3S_1)$&$D^*\bar{B}^*(^3D_1)$&$D^*\bar{B}^*(^5D_1)$
&     -          &     -           &     -   \\
       ~         &  $1(2^+)$   &$D^*\bar{B}^*(^5S_2)$
&$D^*\bar{B}^*(^1D_2)$&$D^*\bar{B}^*(^5D_2)$  &     -           &      -     & -                &     -           &     -    &     -           &      -   \\
\hline
\multirow{3}*{$1$}&${1\over 2}(0^+)$&$D\bar{B}_s(^1S_0)$   &$D_s\bar{B}(^1S_0)$&$D^*\bar{B}_s^*(^1S_0)$&$D_s^*\bar{B}^*(^1S_0)$ &$D^*\bar{B}_s^*(^5D_0)$ &$D_s^*\bar{B}^*(^5D_0)$  &   -        &     -      &
        -        &      -            \\
                 &${1\over 2}(1^+)$&$D_s\bar{B}^*(^3S_1)$
&$D\bar{B}_s^*(^3S_1)$ &$D_s^*\bar{B}(^3S_1)$
&$D^*\bar{B}_s(^3S_1)$
&$D_s^*\bar{B}^*(^3S_1)$&$D^*\bar{B}_s^*(^3S_1)$
&$D_s\bar{B}^*(^3D_1)$
&$D\bar{B}_s^*(^3D_1)$ &$D_s^*\bar{B}(^3D_1)$  &$D^*\bar{B}_s(^3D_1)$ \\
                       &${1\over 2}(2^+)$&$D_s^*\bar{B}^*(^5S_2)$
&$D^*\bar{B}_s^*(^5S_2)$&$D_s^*\bar{B}^*(^1D_2)$&$D^*\bar{B}_s^*(^1D_2)$
&$D_s^*\bar{B}^*(^5D_2)$&$D^*\bar{B}_s^*(^5D_2)$&       -      &     -      &
      -          &     -     \\
\hline
\multirow{3}*{$2$}&$0(0^+)$    &
$D_s\bar{B}_s(^1S_0)$ &$D_s^*\bar{B}_s^*(^1S_0)$&
$D_s^*\bar{B}_s^*(^5D_0)$&    -    &    -      &    -      &   -   &   -    &
   -             &    -   \\
        ~        &  $0(1^+)$    &$D_s\bar{B}_s^*(^3S_1)$&$D_s\bar{B}_s^*(^3D_1)$ &
$D_s^*\bar{B}_s(^3S_1)$&$D_s^*\bar{B}_s(^3D_1)$&$D_s^*\bar{B}_s^*(^3S_1)$
&$D_s^*\bar{B}_s^*(^3D_1)$&$D_s^*\bar{B}_s^*(^5D_1)$&   -  &   -   &   -   \\
       ~         & $0(2^+)$   &$D_s^*\bar{B}_s^*(^5S_2)$
&$D_s^*\bar{B}_s^*(^1D_2)$&$D_s^*\bar{B}_s^*(^5D_2)$&   -   &  -   &   -   &
       -         &     -  &   -    &    -     \\
\bottomrule[1.0pt]
\end{tabular*}
\end{table*}

The Lagrangians under the heavy quark symmetry and the
SU(3)-flavor symmetry
read~\cite{PhysRevD.45.R2188,Falk:1992cx,Grinstein:1992qt,Casalbuoni:1996pg}
\begin{eqnarray}
\mathcal{L}_{HH\mathcal{M}}&=& ig\mbox{Tr}\left[H^{(Q)}_b \gamma_\mu\gamma_5
A_{ba}^\mu \bar{H}^{(Q)}_a\right],
\label{lagrangian:p}\\
\mathcal{L}_{HHV}&=& i\beta\mbox{Tr}\left[ H^{(Q)}_b v_\mu
(V^\mu_{ba}-\rho^\mu_{ba})\bar{H}^{(Q)}_a\right]\nonumber\\
&&+i\lambda\mbox{Tr}\left[ H^{(Q)}_b\sigma_{\mu\nu}F^{\mu\nu}(\rho)_{ba}\bar{H}^{(Q)}_a\right], \\
\mathcal{L}_{ HH\sigma}&=&g_s \mbox{Tr}\left[H^{(Q)}_a\sigma
\bar{H}^{(Q)}_a\right], \label{lagrangian:v}
\end{eqnarray}
where $H^{(Q)}$ and $\bar{H}^{(Q)}$ are defined as
\begin{eqnarray}
    H_a^{(Q)}&=&\frac{1+\not{v}}{2}[{P}^{*\mu}_{a}\gamma_{\mu}
    -{P}_a\gamma_5]
\end{eqnarray}
and
\begin{eqnarray}
    \bar{H}_a^{(Q)}\equiv\gamma_0H^{(Q)\dag}\gamma_0=[{P}^{*\dag\mu}_{a}\gamma_{\mu}
    +{P}^{\dag}_a\gamma_5]\frac{1+\not{v}}{2}
\end{eqnarray}
with $P^{*}_a=(D^{*0}, D^{*+},D_s^{*+})$ or $(B^{*-},
\bar{B}^{*0},\bar{B}_s^{*0})$ being the charmed or antibottomed
vector mesons and $P=(D^{0}, D^{+},D_s^{+})$ or $(B^{-},
\bar{B}^{0},\bar{B}_s^{0})$ being the charmed or anti-bottomed
pseudoscalar mesons. The trace acts on the gamma matrices. The
axial-current $A^{\mu}$ is defined as $A^{\mu}\equiv{1\over
2}\left(\xi^{\dag}\partial^{\mu}\xi-\xi\partial^{\mu}\xi^{\dag}\right)={i\over
f_{\pi}}\partial^{\mu}\mathcal{M}+\ldots$ where $\xi\equiv e^{iM\over
f_{\pi}}$ with $\mathcal{M}$ being the exchanged pseudoscalar meson matrix
given in Eq.~(\ref{exchangeP}). The vector current $V^{\mu}$ is
defined as $V^{\mu}\equiv{1\over
2}\left(\xi^{\dag}\partial^{\mu}\xi+\xi\partial^{\mu}\xi^{\dag}\right)$.
In the heavy quark limit, the heavy meson velocity is adopted as
$v^{\mu}=(1,0,0,0)$.
$F_{\mu\nu}(\rho)\equiv\partial_{\mu}\rho_{\nu}-\partial_{\nu}\rho_{\mu}-
\left[\rho_{\mu},\rho_{\nu}\right]$ where $\rho_{\mu}={ig_{v}\over
\sqrt{2}}\hat{\rho}_{\mu}$ with $\hat{\rho}$ being the exchanged
vector meson matrix given in Eq.~\ref{exchangeV}. Expanding the
Lagrangians given in Eqs.~(\ref{lagrangian:p}-\ref{lagrangian:v}),
we list the specific expressions in
Eqs.~(\ref{Lagrangian-PvPvM}-\ref{Lagrangian-PvPvs}).

\begin{eqnarray}
    \mathcal{M}&=&\left(\begin{array}{ccc}
        \frac{1}{\sqrt{2}}\pi^0+\frac{\eta}{\sqrt{6}}&\pi^+&K^+\\
        \pi^-&-\frac{1}{\sqrt{2}}\pi^0+\frac{\eta}{\sqrt{6}}&K^0\\
        K^-  &\bar{K}^0&-\frac{2}{\sqrt{6}}\eta\\
\end{array}\right),\label{exchangeP}
\end{eqnarray}
\begin{eqnarray}
\hat{\rho}^{\mu}&=&\left(\begin{array}{ccc}
\frac{\rho^{0}}{\sqrt{2}}+\frac{\omega}{\sqrt{2}}&\rho^{+}&K^{*+}\\
\rho^{-}&-\frac{\rho^{0}}{\sqrt{2}}+\frac{\omega}{\sqrt{2}}&K^{*0}\\
K^{*-}&\bar{K}^{*0}&\phi\\
\end{array}\right)^{\mu}.\label{exchangeV}
\end{eqnarray}

\begin{eqnarray}
\mathcal{L}_{P^*P^*\mathcal{M}} &=&
-i\frac{2g}{f_\pi}\varepsilon_{\alpha\mu\nu\lambda} v^\alpha
P^{*\mu}_{b}{P}^{*\lambda\dag}_{a}
\partial^\nu{}M_{ba},\label{Lagrangian-PvPvM}\\
\mathcal{L}_{P^*P\mathcal{M}} &=&- \frac{2g}{f_\pi}\left(P^{}_b
P^{*\dag}_{a\lambda}+
P^{*}_{b\lambda}P^{\dag}_{a}\right)\partial^\lambda{}
M_{ba},\label{Lagrangian-PvPM}
\end{eqnarray}
\begin{eqnarray}
  \mathcal{L}_{PPV}
  &=& -\sqrt{2}\beta{}g_V P_bP_a^{\dag}
  v\cdot\hat{\rho}_{ba},\label{Lagrangian-PPV}\\
  \mathcal{L}_{P^*PV}
  &=&- 2\sqrt{2}\lambda{}g_V v^\lambda\varepsilon_{\lambda\mu\alpha\beta}
  \left(P_bP^{*\mu\dag}_a +P_b^{*\mu}P^{\dag}_a\right)
  \left(\partial^\alpha{}\hat{\rho}^\beta\right)_{ba},\label{Lagrangian-PvPV}\\
  \mathcal{L}_{P^*P^*V}
  &=& \sqrt{2}\beta{}g_V P_b^{*}\cdot P^{*\dag}_a
  v\cdot\hat{\rho}_{ba}\nonumber\\
  &&-i2\sqrt{2}\lambda{}g_V P^{*\mu}_b P^{*\nu\dag}_a
  \left(\partial_\mu{}
  \hat{\rho}_\nu - \partial_\nu{}\hat{\rho}_\mu\right)_{ba},\label{Lagrangian-PvPvV}
  \end{eqnarray}
 \begin{eqnarray}
  \mathcal{L}_{PP\sigma}
  &=& -2g_s P^{}_b P^{\dag}_b\sigma, \label{Lagrangian-PPs} \\
  \mathcal{L}_{P^*P^*\sigma}
  &=& 2g_sP^{*}_b\cdot P^{*\dag}_b\sigma,\label{Lagrangian-PvPvs}
\end{eqnarray}

In the above, $f_{\pi}=132$ MeV is the pion decay constant. The
coupling constant $g$ was studied by many theoretical approaches,
such as quark model~\cite{Falk:1992cx} and QCD sum
rule~\cite{Dai:1998vh,Navarra:2000ji}. In our study, we take the
experimental result of the CLEO Collaboration,
$g=0.59\pm0.07\pm0.01$, which was extracted from the full width of
$D^{*+}$ ~\cite{Ahmed:2001xc}. For the coupling constants relative
to the vector meson exchange, we adopt the values $g_v=5.8$ and
$\beta=0.9$ which were determined by the vector meson dominance
mechanism, and $\lambda=0.56~\mbox{GeV}^{-1}$ which was obtained
by matching the form factor predicted by the effective theory
approach with that obtained by the light cone sum rule and the
lattice QCD simulation~\cite{Isola:2003fh,Bando:1987br}. The
coupling constant for the scalar meson exchange is
$g_s=g_{\pi}/(2\sqrt{6})$~\cite{Liu2008a} with $g_{\pi}=3.73$. We
take the masses of the heavy mesons and the exchanged light mesons
from PDG~\cite{0954-3899-37-7A-075021} and summarize them in
Table~\ref{Mass}.

\begin{table}[htp]
\renewcommand{\arraystretch}{1.0}
\caption{The masses of the heavy mesons and the exchanged light
mesons taken from the PDG~\cite{0954-3899-37-7A-075021}. In our
study, we keep the isospin symmetry. For the isospin multiplet, we
use the averaged mass in our study.}\label{Mass}
\begin{tabular*}{8.5cm}{@{\extracolsep{\fill}}cccc}
\toprule[0.8pt]  \addlinespace[2pt]
Heavy Mesons    &  Mass (MeV)   &  Exchanged Mesons   &   Mass (MeV)  \\
\hline
$D^{\pm}$       & $1869.60$     &  $\pi^{\pm}$        & $139.57$     \\
$D^0$           & $1864.83$     &  $\pi^0$            & $134.98$     \\
$D^{*\pm}$      & $2010.25$     &  $\eta$             & $547.85$     \\
$D^{*0}$        & $2006.96$     &  $\rho$             & $775.49$     \\
$D_s^{\pm}$     & $1968.47$     &  $\omega$           & $782.65$     \\
$D_s^{*\pm}$    & $2112.3$      &  $\phi$             & $1019.46$    \\
$B^*$           & $5325.1$      &  $\sigma$           & $600$        \\
$B^{\pm}$       & $5279.17$     &  $K^{\pm}$          & $493.67$     \\
$B^0$           & $5279.50$     &  $K^0$              & $497.61$     \\
$B_s^0$         & $5366.3$      &  $K^{*\pm}$         & $891.66$     \\
$B_s^*$         & $5415.4$      &  $K^{*0}$           & $895.94$     \\
\hline
\end{tabular*}
\end{table}

\subsection{The derivation of the effective potentials}\label{potential}
Using the lagrangians given in
Eqs.~(\ref{Lagrangian-PvPvM}-\ref{Lagrangian-PvPvs}), one can
easily deduce the effective potentials in the momentum space. 
Taking into account the structure effect of the heavy mesons, we
introduce a monopole form factor
\begin{eqnarray}
F(q)={\Lambda^2-m_{ex}^2 \over \Lambda^2-q^2}
\end{eqnarray}
at each vertex. Here, $\Lambda$ is the cutoff parameter and
$m_{ex}$ is the mass of the exchanged meson. 

We need to emphasize that there is alternative approach to deduce the effective potential just shown in Refs. \cite{Cordon:2009pj,Gamermann:2009uq}, where they refuse to introduce the form factor  due to the lack of knowledge of form factors. Here, one can also regularize the divergence of the potential at short distance from a renormalization viewpoint. For the detailed information of the renormalization approach,  the interested readers can refer to Refs. \cite{Cordon:2009pj,Gamermann:2009uq}, where the coordinate space renormalization, i.e., boundary conditions, is adopted.

Making fourier
transformation
\begin{eqnarray}
V(r)={1\over (2\pi)^3}\int d{\bf q}^3 e^{-i{\bf q}\cdot{\bf
r}}V({\bf q})F^2({\bf q})
\end{eqnarray}
one can obtain the effective potentials in the coordinate space.
In  Eqs.~(\ref{PPpotential}-\ref{VVpotential}), we list the
specific expressions of the effective subpotentials which are
flavor-independent. The effective potential used in our
calculation is the product of the flavor-independent subpotentials
and the isospin-dependent coefficients which are summarized in the
Appendix~\ref{Appendix}. The flavor-independent subpotentials are

\begin{eqnarray}
V_{\sigma}^a(r) &=& -g_s^2H_0(\Lambda,q_0,m_\sigma,r),\\
V_{v}^a(r) &=&
\frac{\beta^2g_V^2}{2}H_0(\Lambda,q_0,m_v,r),\label{PPpotential}
\end{eqnarray}
for the process $PP\to PP$,
\begin{eqnarray}
V_{p/\sigma/v}^b(r) &=& 0
\end{eqnarray}
for the scattering process $PP \to PP^*$,
\begin{eqnarray}
V_{p}^c (r) &=& \frac{g^2}{f_\pi^2}\Big[H_3(\Lambda,q_0,m_{p},r)T(\mbox{\boldmath $\epsilon$}_3^{\dag},\mbox{\boldmath$\epsilon$}_4^{\dag})\nonumber\\
&~&+\frac{1}{3}H_1(\Lambda,q_0,m_{p},r)S(\mbox{\boldmath$\epsilon$}_3^{\dag},
\mbox{\boldmath$\epsilon$}_4^{\dag})\Big]\\
V_{v}^c(r) &=& -2\lambda^2g_V^2\Big[H_3(\Lambda,q_0,m_{v},r)T(\mbox{\boldmath $\epsilon$}_3^{\dag},\mbox{\boldmath$\epsilon$}_4^{\dag})\nonumber\\
&&-\frac{2}{3}H_1(\Lambda,q_0,m_{v},r)S(\mbox{\boldmath$\epsilon$}_3^{\dag},
\mbox{\boldmath$\epsilon$}_4^{\dag})\Big]
\end{eqnarray}
for the scattering process $PP\to P^*P^*$,
\begin{eqnarray}
V_{\sigma}^d(r) &=&
-g_s^2H_0(\Lambda,q_0,m_{\sigma},r)S(\mbox{\boldmath$\epsilon$}_4^{\dag},
\mbox{\boldmath$\epsilon$}_2) \\
V_{v}^d(r) &=& \frac{\beta^2g_V^2}{2}H_0(\Lambda,q_0,m_{v},r)
S(\mbox{\boldmath$\epsilon$}_4^{\dag},\mbox{\boldmath$\epsilon$}_2)
\end{eqnarray}
for the scattering process $PP^*\to PP^*$,
\begin{eqnarray}
V_{\pi}^e (r) &=&
\frac{g^2}{f_{\pi}^2}\Big[M_3(\Lambda,q_0,m_{\pi},r)
T(\mbox{\boldmath$\epsilon$}_3^{\dag},\mbox{\boldmath$\epsilon$}_2) \nonumber\\
&~&+\frac{1}{3}M_1(\Lambda,q_0,m_{\pi},r)S(\mbox{\boldmath$\epsilon$}_3^{\dag},\mbox{\boldmath$\epsilon$}_2)\Big]
\label{PPv-PvP-pi}\\
V_{\pi/\eta/K}^e (r) &=&
\frac{g^2}{f_\pi^2}\Big[H_3(\Lambda,q_0,m_{\pi/\eta/K},r)
T(\mbox{\boldmath$\epsilon$}_3^{\dag},\mbox{\boldmath$\epsilon$}_2) \nonumber\\
&~&+\frac{1}{3}H_1(\Lambda,q_0,m_{\pi/\eta/K},r)S(\mbox{\boldmath$\epsilon$}_3^{\dag},\mbox{\boldmath$\epsilon$}_2)\Big]\\
V_{v}^e(r) &=&
-2\lambda^2g_V^2\Big[H_3(\Lambda,q_0,m_{v},r)T(\mbox{\boldmath$\epsilon$}_3^{\dag},
\mbox{\boldmath$\epsilon$}_2)\nonumber\\
&~&-\frac{2}{3}H_1(\Lambda,q_0,m_{v},r)S(\mbox{\boldmath$\epsilon$}_3^{\dag},
\mbox{\boldmath$\epsilon$}_2)\Big]
\end{eqnarray}
for the scattering process $PP^*\to P^*P$,
\begin{eqnarray}
V_{p}^f (r) &=&\frac{g^2}{f_\pi^2}\Big[H_3(\Lambda,q_0,m_{p},r)
T(\mbox{\boldmath$\epsilon$}_3^{\dag},i\mbox{\boldmath$\epsilon$}_4^{\dag}\times\mbox{\boldmath$\epsilon$}_2)
\nonumber\\
&~&+\frac{1}{3}H_1(\Lambda,q_0,m_{p},r)T(\mbox{\boldmath$\epsilon$}_3^{\dag},
i\mbox{\boldmath$\epsilon$}_4^{\dag}\times\mbox{\boldmath$\epsilon$}_2)\Big]\label{PP-PVPV-pseudo}\\
V_{v}^f (r) &=& 2\lambda^2g_V^2\Big\{H_3(\Lambda,q_0,m_{v},r) \nonumber \\
&~&\times\big[T(i\mbox{\boldmath$\epsilon$}_3^{\dag}\times\mbox{\boldmath$\epsilon$}_4^{\dag},
\mbox{\boldmath$\epsilon$}_2)-T(i\mbox{\boldmath$\epsilon$}_3^{\dag}\times\mbox{\boldmath$\epsilon$}_2^{\dag},
\mbox{\boldmath$\epsilon$}_4^{\dag})\big]\nonumber \\
&~&+\frac{1}{3}H_1(\Lambda,q_0,m_{v},r)\nonumber\\
&~&\times\big[S(i\mbox{\boldmath$\epsilon$}_3^{\dag}\times\mbox{\boldmath$\epsilon$}_4^{\dag},
\mbox{\boldmath$\epsilon$}_2)-S(i\mbox{\boldmath$\epsilon$}_3^{\dag}\times\mbox{\boldmath$\epsilon$}_2^{\dag},
\mbox{\boldmath$\epsilon$}_4^{\dag})\big]\Big\}\label{PP-PVPV-vector}
\end{eqnarray}
for the scattering process $PP^*\to P^*P^*$, and
\begin{eqnarray}
V_{p}^h (r) &=&\frac{g^2}{f_{\pi}^2}\Big[H_3(\Lambda,q_0,m_{p},r)
T\left(i\mbox{\boldmath$\epsilon$}_3^{\dag}\times\mbox{\boldmath$\epsilon$}_1,
i\mbox{\boldmath$\epsilon$}_4^{\dag}\times\mbox{\boldmath$\epsilon$}_2\right) \nonumber\\
&~&+\frac{1}{3}H_1(\Lambda,q_0,m_{p},r)S\left(i\mbox{\boldmath$\epsilon$}_3^{\dag}\times\mbox{\boldmath$\epsilon$}_1,
i\mbox{\boldmath$\epsilon$}_4^{\dag}\times\mbox{\boldmath$\epsilon$}_2\right)\Big]\\
V_{\sigma}^h (r) &=& -g_s^2H_0(\Lambda,q_0,m_{\sigma},r)
C\left(i\mbox{\boldmath$\epsilon$}_3^{\dag}\times\mbox{\boldmath$\epsilon$}_1,
i\mbox{\boldmath$\epsilon$}_4^{\dag}\times\mbox{\boldmath$\epsilon$}_2\right) \\
V_{v}^h (r) &=& \frac{\beta^2g_V^2}{2}H_0(\Lambda,q_0,m_{v},r)
C\left(i\mbox{\boldmath$\epsilon$}_3^{\dag}\times\mbox{\boldmath$\epsilon$}_1,
i\mbox{\boldmath$\epsilon$}_4^{\dag}\times\mbox{\boldmath$\epsilon$}_2\right)\nonumber\\
&~&-2\lambda^2g_V^2\Big[H_3(\Lambda,q_0,m_{v},r)
T\left(i\mbox{\boldmath$\epsilon$}_3^{\dag}\times\mbox{\boldmath$\epsilon$}_1,
i\mbox{\boldmath$\epsilon$}_4^{\dag}\times\mbox{\boldmath$\epsilon$}_2\right)\nonumber\\
&~&-\frac{2}{3}H_1(\Lambda,q_0,m_{v},r)
S\left(i\mbox{\boldmath$\epsilon$}_3^{\dag}\times\mbox{\boldmath$\epsilon$}_1,
i\mbox{\boldmath$\epsilon$}_4^{\dag}\times\mbox{\boldmath$\epsilon$}_2\right)\Big]\label{VVpotential}
\end{eqnarray}
for the scattering process $P^*P^*\to P^*P^*$. To obtain the
effective potentials $V^{g}(r)$ for the process $P^*P\to P^*P^*$,
one just needs to make the following changes \begin{eqnarray}
\mbox{\boldmath$\epsilon$}_3^{\dag}\to\mbox{\boldmath$\epsilon$}_4^{\dag},&\quad&
i\mbox{\boldmath$\epsilon$}_4^{\dag}\times\mbox{\boldmath$\epsilon$}_2\to i\mbox{\boldmath$\epsilon$}_3^+\times\mbox{\boldmath$\epsilon$}_1^{\dag},\nonumber\\
\mbox{\boldmath$\epsilon$}_2\to\mbox{\boldmath$\epsilon$}_1,&\quad&
i\mbox{\boldmath$\epsilon$}_3^{\dag}\times\mbox{\boldmath$\epsilon$}_4^{\dag}\to i\mbox{\boldmath$\epsilon$}_4^{\dag}\times\mbox{\boldmath$\epsilon$}_3^{\dag},\\
\mbox{\boldmath$\epsilon$}_4^{\dag}\to\mbox{\boldmath$\epsilon$}_3^{\dag},&\quad&
i\mbox{\boldmath$\epsilon$}_3^{\dag}\times\mbox{\boldmath$\epsilon$}_2\to
i\mbox{\boldmath$\epsilon$}_4^{\dag}\times\mbox{\boldmath$\epsilon$}_1,\nonumber
\end{eqnarray}
in Eqs.(~\ref{PP-PVPV-pseudo})-(\ref{PP-PVPV-vector}). Functions
$H_0$, $H_1$, $H_3$, $M_1$ and $M_3$ are given in the Appendix.
Operator $C$, the generalized tensor operator $T$ and spin-spin
operator $S$ are defined as
\begin{eqnarray}
 C(a,b)&=&a b, \\
T(\mbox{\boldmath$a$},\mbox{\boldmath$b$})&=&
\frac{3\mbox{\boldmath$a$}\cdot\mbox{\boldmath$r$}\mbox{\boldmath$b$}\cdot\mbox{\boldmath$r$}}{r^2}-
\mbox{\boldmath$a$}\cdot\mbox{\boldmath$b$},\\
S(\mbox{\boldmath$a$},\mbox{\boldmath$b$})&=&\mbox{\boldmath$a$}\cdot\mbox{\boldmath$b$}.
\end{eqnarray}

Due to the large mass gap between the mesons $D(D^0,D^+)$,
$D_s^+$,  $D^*(D^{*0},D^{*+})$ and $D_s^{*+}$ (similarly, in the
bottom sector), it is necessary to adopt the nonzero time
component of the transferred momentum for some scattering
processes. We present the $q_0$s used in our calculation in the
Appendix. Notice that $m_{D^*}-m_{D}>m_{\pi}$ leads to the complex
potential for the scattering process $DD^*\to D^*D$, and we take
its real part which has an oscillation form, see
Eq.~(\ref{PPv-PvP-pi}).

\section{Numerical Results}\label{Results}
Using the potentials given in the subsection~\ref{potential}, we
solve the coupled-channel Schr\"odinger equation and summarize the
numerical results which include the binding energy (B.E.), the
system mass (M), the root-mean-square radius ($r_{rms}$) and the
probability of the individual channel ($P_i$) in
Tables~\ref{numerical:DD},~\ref{numerical:BB},
\ref{numerical:DDBBs},~\ref{numerical:DBs},~\ref{numerical:DDBBss}and
\ref{numerical:DBss}.

In our study, only the cutoff is a
free parameter. However, due to the lack of the experimental data,
one can not determine the cutoff exactly. Thus, it is very
difficult to draw definite conclusions. Luckily the one-boson-exchange potential model is very successful to describe the deuteron with the cutoff in the range $0.8<\Lambda<1.5$ GeV. Following the study of the deuteron with the same formalism and taking into account the mass difference between the heavy meson and the nucleon, we take the range of the cutoff to be
$0.9~\mbox{GeV}<\Lambda<2.5~\mbox{GeV}$. However, this choice is a
little arbitrary to some extent. We sincerely hope that in the
near future there will be enough experimental data with which one
can determine the cutoff exactly.

Besides, we also consider the stability of the results when we draw our conclusions.

\subsection{The Numerical Results for Systems with Strangeness $S=0$}

For the systems with strangeness $S=0$, in order to highlight the
role of the long-range pion exchange in the formation of the
loosely bound state, we first give the numerical results with the
pion-exchange potential alone, which are marked with OPE, and then
with the heavier eta, sigma, rho and omega exchanges as well as
the pion exchange, which are marked with OBE, see
Tables~\ref{numerical:DD},~\ref{numerical:BB}
and~\ref{numerical:DB}.

\subsubsection{$D^{(*)}D^{(*)}$}

\begin{figure}[htp]
\begin{tabular}{cc}
\includegraphics[width=0.24\textwidth]{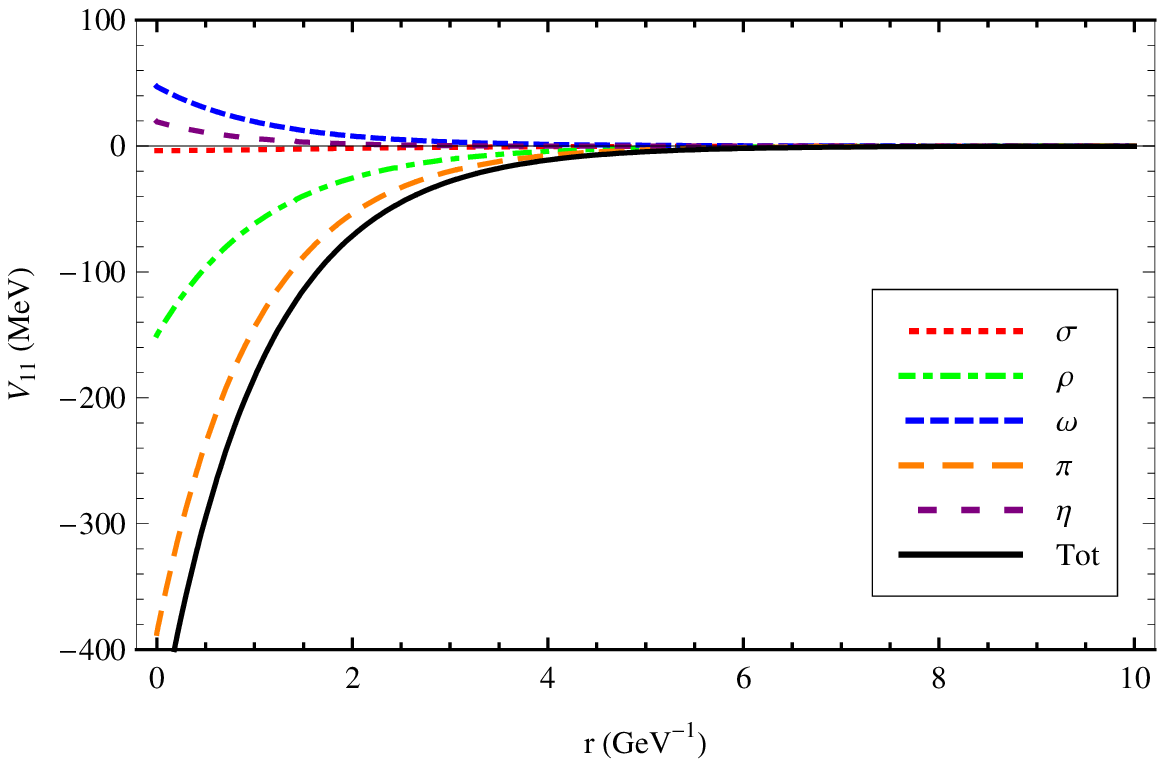} &
\includegraphics[width=0.24\textwidth]{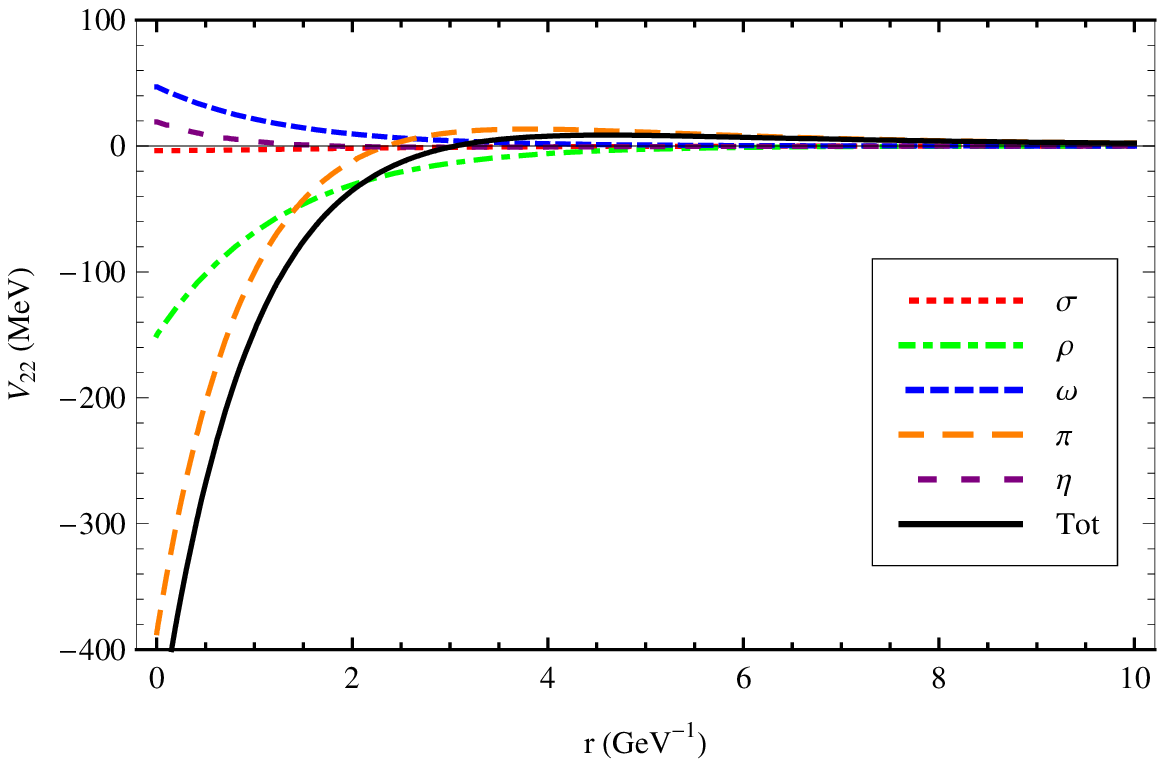} \\
$(V_{11})$&$(V_{22})$\\
\includegraphics[width=0.24\textwidth]{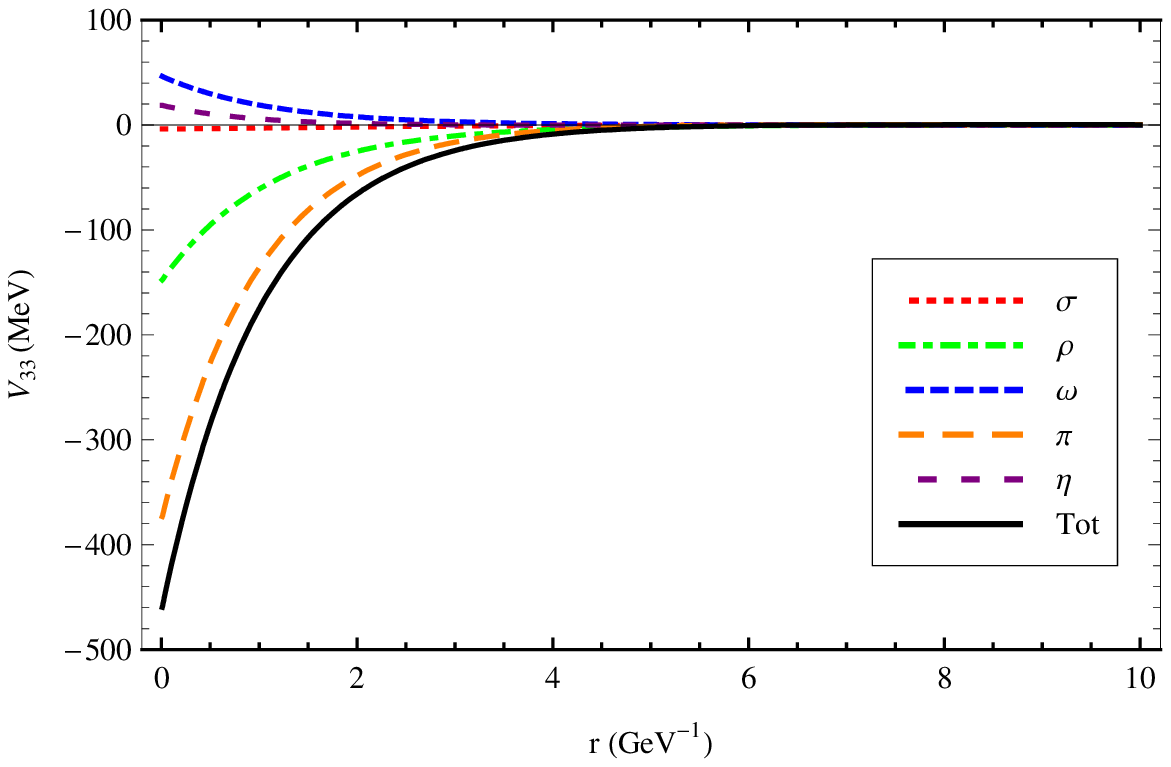} &
\includegraphics[width=0.24\textwidth]{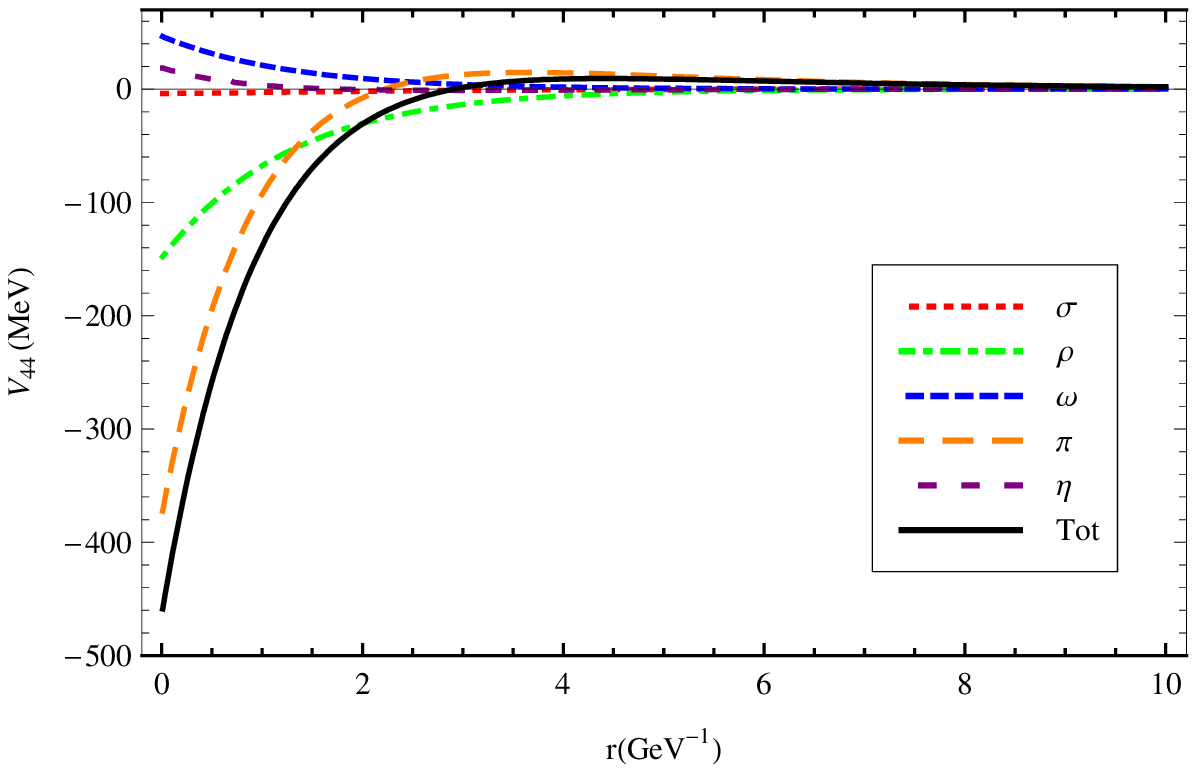} \\
$(V_{33})$&$(V_{44})$ \\
\includegraphics[width=0.24\textwidth]{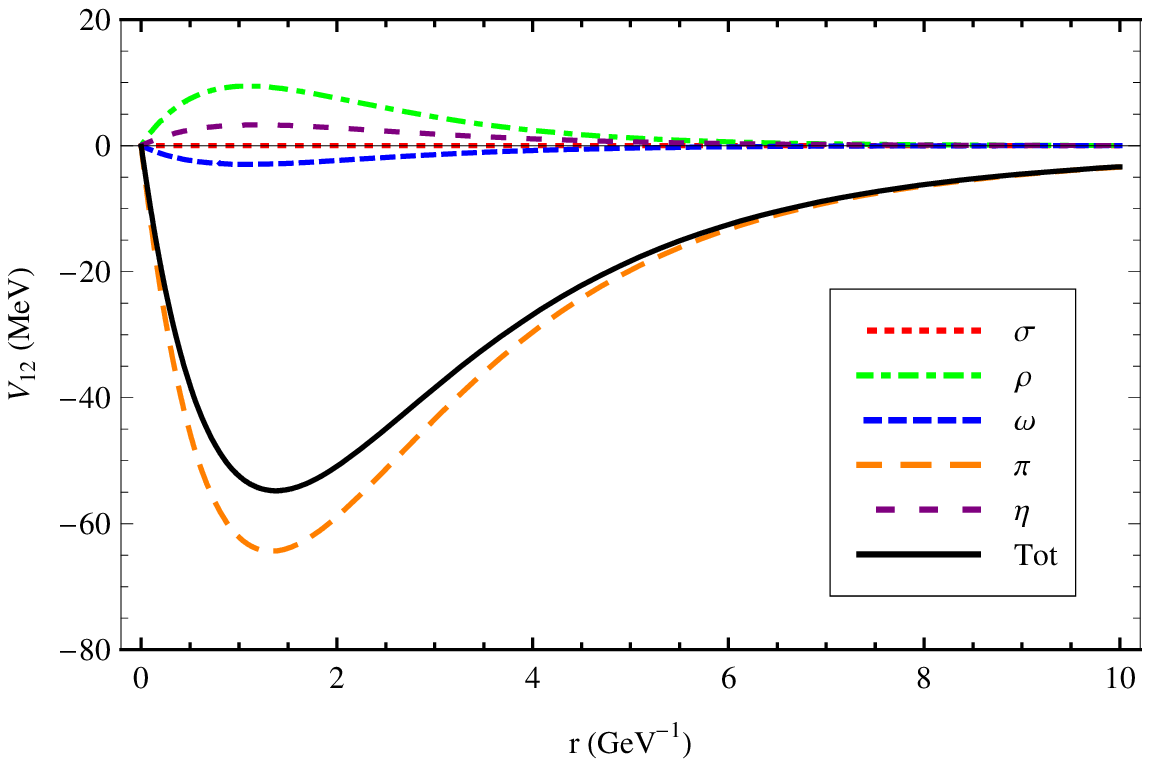} &
\includegraphics[width=0.24\textwidth]{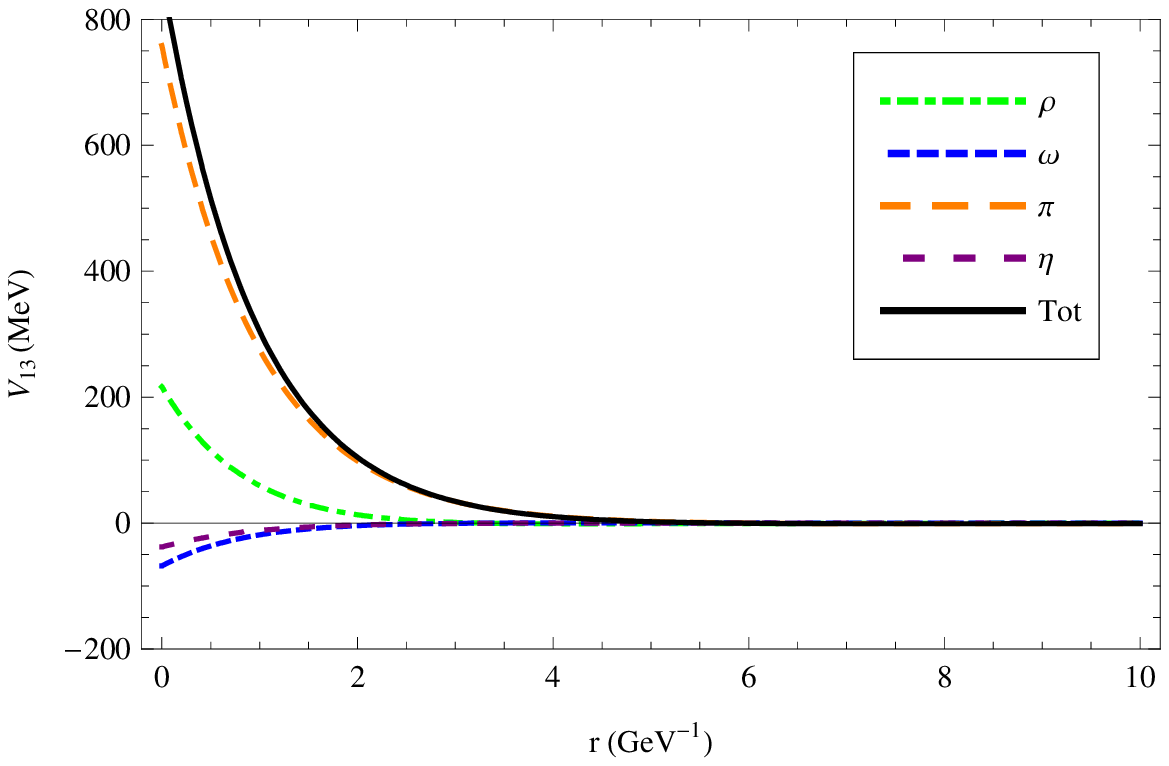} \\
$(V_{12})$&$(V_{13})$\\
\includegraphics[width=0.24\textwidth]{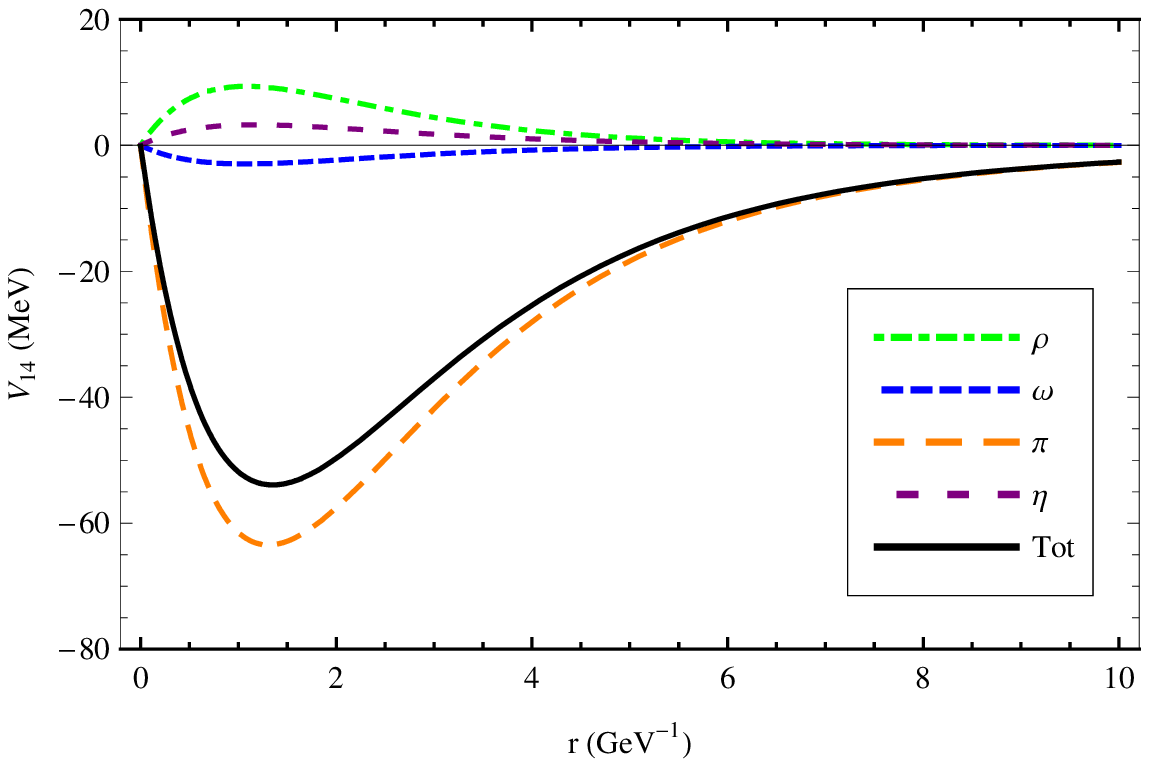}&
\includegraphics[width=0.24\textwidth]{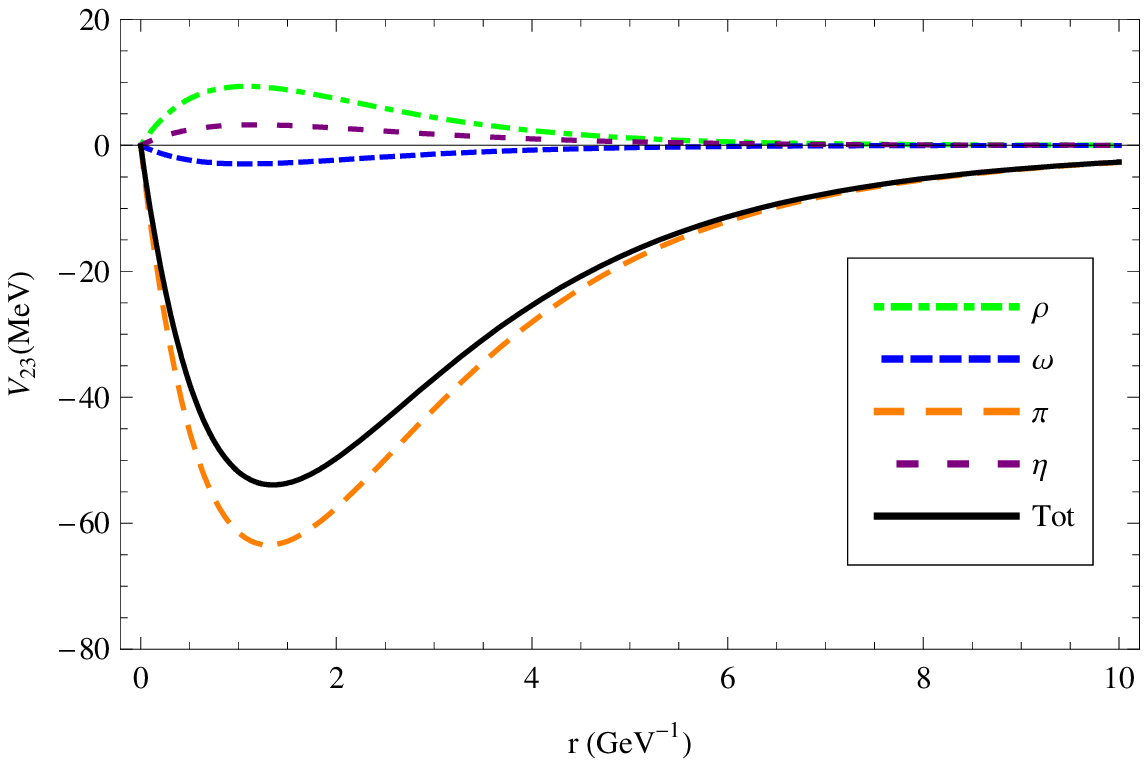}\\
$(V_{14})$&$(V_{23})$\\
\includegraphics[width=0.24\textwidth]{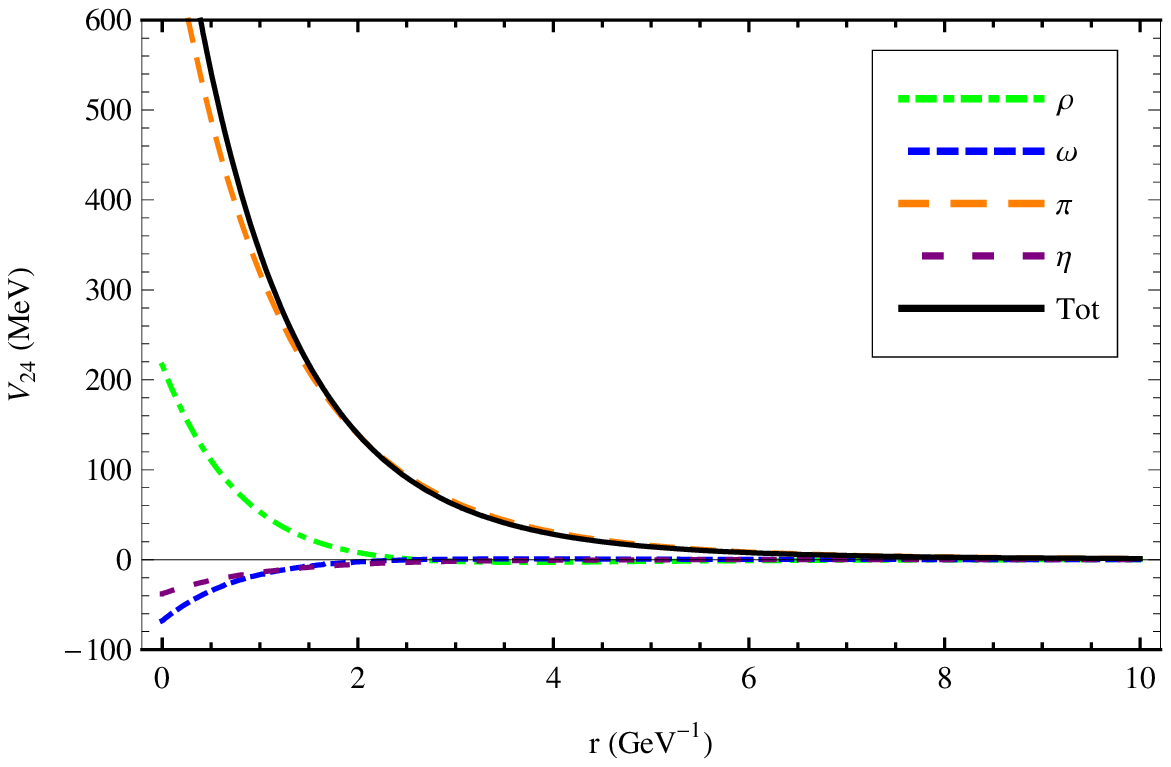}&
\includegraphics[width=0.24\textwidth]{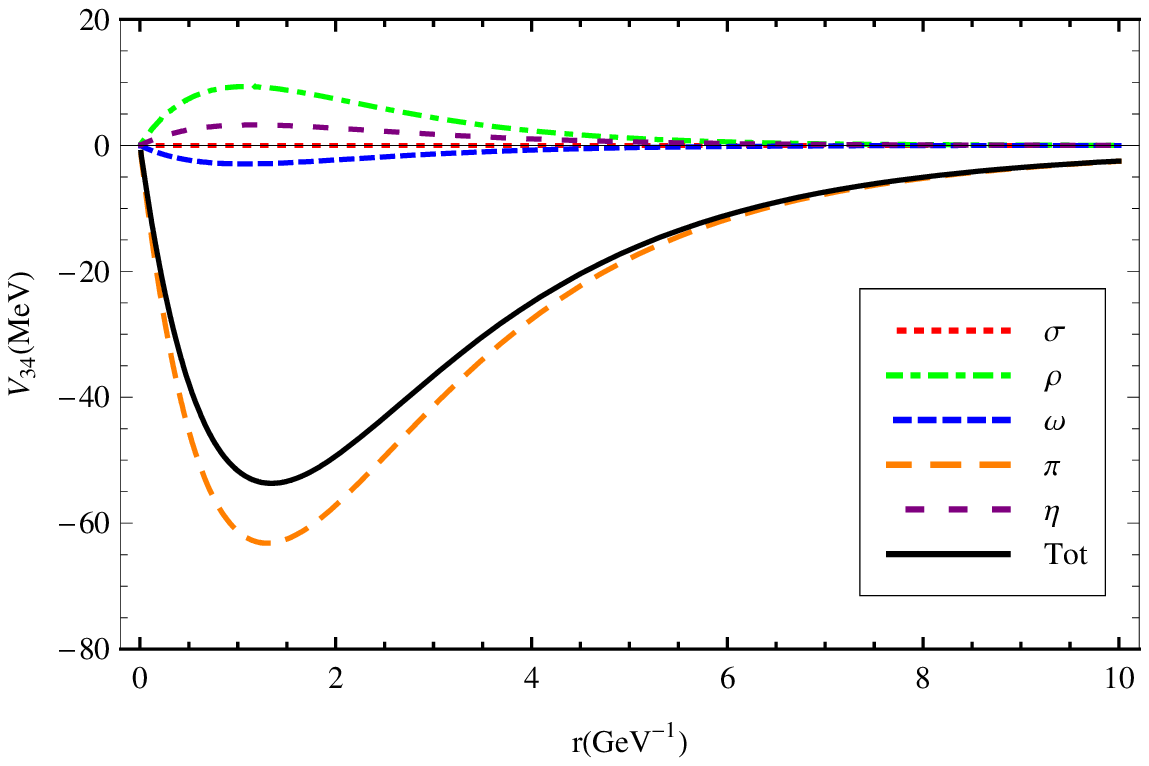}\\
$(V_{24})$&$(V_{34})$\\
\end{tabular}
\caption{(Color online). The effective potentials for the state
$D^{(*)}D^{(*)}[I(J^P)=0(1^+)]$ with $\mbox{cutoff}=1.00$
GeV.}\label{plot:potential}
\end{figure}

\begin{figure}
  \centering
  \includegraphics[width=0.48\textwidth]{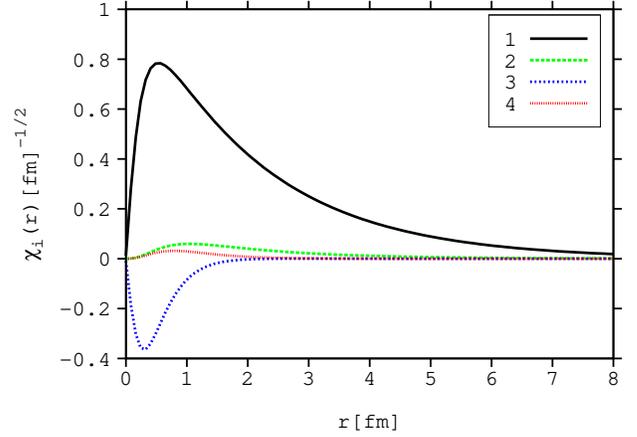}
  \caption{(Color online). The wave function (r$\psi(r)$)of the $D^{(*)}D^{(*)}$ system with strangeness 0 and $I(J^P)=0(1^+)$ when
  the cutoff parameter is fixed to be 1.0 GeV.}\label{plot:wfunction}
\end{figure}

The state $D^{(*)}D^{(*)}[I(J^P)=0(0^+)]$ is forbidden because the
present boson system should satisfy the Boson-Einstein statistic.
However, the state $D^{(*)}D^{(*)}[I(J^P)=0(1^+)]$ is very
interesting. Using the long-range pion exchange potential, we obtain
a loosely bound state with a reasonable cutoff. For our present
$D^{(*)}D^{(*)}[I(J^P)=0(1^+)]$ state, with the cutoff parameter
fixed larger than $1.05$ GeV, the long-range pion exchange is strong
enough to form the loosely bound state. If we set the cutoff
parameter to be $1.05$ GeV, the binding energy relative to the
$DD^*$ threshold is $1.24$ MeV and the corresponding
root-mean-square radius is $3.11$ fm which is comparable to the size
of the deuteron (about $2.0$ fm). The dominant channel is
$[DD^*]_{-}(^3S_1)$, with a probability $96.39\%$. With such a large
mass gap (about $140$ MeV) between the threshold of $DD^*$ and that
of $D^*D^*$, the contribution of the state $D^*D^*(^3S_1)$ is
$2.79\%$. However, the probability of the D-wave is around $1\%$.
When we tune the cutoff to be $1.20$ GeV, the binding energy is
$20.98$ MeV and the root-mean-square radius changes into $0.84$ fm.
When we use the one-boson-exchange potential, we notice that the
binding becomes deeper. For example, if the cutoff is fixed at
$1.10$ GeV, the binding energy is $4.63$ MeV with OPE potential.
However, it changes into $42.82$ GeV with the OBE potential for the
same cutoff, see Table~\ref{numerical:DD}. We also plot the
potentials in Fig.~\ref{plot:potential}. From the potentials, one
can see that the heavier rho and omega exchanges cancel each other
significantly, which can be easily understood since
for the isospin-zero system the isospin factor of $\rho$ is $-3$
while that of $\omega$ is 1. From the the potentials $V_{11}$,
$V_{22}$, $V_{33}$ and $V_{44}$ of Fig.~\ref{plot:potential}, one
can also see clearly that the total potential is below the
$\pi$-exchange potential and the contributions of the $\eta$ and
$\sigma$ exchanges are very small. This implies that the total
potential of the $\rho$ and $\omega$ exchanges is helpful to
strengthen the binding. 
We note that the OBE potentials deduced by introducing the form factor generate spurious deeply bound states \cite{Cordon:2009pj}. In order to fix this problem, we also plot the wave function in
Fig.~\ref{plot:wfunction} from which one can see that there is no
node except the origin. In other words, it is really a ground
state.

To see the effect of the $\sigma$, $\rho$ and
$\omega$ exchanges, we turn off the contributions of the $\pi$ and
$\eta$ exchanges and do the calculation again. We obtain a loosely
bound state with binding energy being 0.78 MeV and root-mean-square
radius being 3.74 fm when the cutoff parameter is fixed to be 1.44
GeV, which is much larger than 1.05 GeV used in the
one-pion-exchange case with almost the same binding energy. Again,
this means that the contribution of the long-range pion exchange is
larger than that of the heavier vector meson exchange in the
formation of the loosely bound $D^{(*)}D^{(*)}[I(J^P)=0(1+)]$ state.
This is different from the conclusion of the
paper~\cite{Nieves:2012tt} in which the authors studied the charmed
meson-charmed anti-meson systems with a effective field theory. In
their power counting, the leading order contribution arises from the
four-meson contact interaction and the one-pion-exchange interaction
is perturbative. The interested reader can refer to the
paper~\cite{Nieves:2012tt} for detailed information. With the
numerical results and the analysis above, the state
$D^{(*)}D^{(*)}[I(J^P)=0(1^+)]$ might be a good molecule candidate.

We should mention that in the calculation of the $X(3872)$ one also obtained a bound $D\bar{D}^*$ state
with quantum numbers $I(J^{PC})=0(1^{++})$ using the OPE potential~\cite{Lee:2009hy,Li:2012cs}. One may be confused since
the difference between the potential of the $DD^*$ system and that
of the $D\bar{D}^*$ system is the G-parity of the exchanged meson
while the pion has an odd G-parity. Actually, the iso-singlet
$D\bar{D}^*$ system has two C-parity states, one with even
C-parity ($C=+$) and the other with odd C-parity ($C=-$). And, the
interaction of our present $DD^*$ system relates to that of the
odd C-parity but not the even C-parity $D\bar{D}^*$ state via the
G-parity rule.

We obtain no binding solutions for the state
$D^{(*)}D^{(*)}[I(J^P)=0(2^+)]$ even if we tune the cutoff
parameter as high as 3.0 GeV. It seems that the present meson-exchange model does not support the state $D^{(*)}D^{(*)}[I(J^P)=0(2^+)]$ to be a molecule.

For the state $D^{(*)}D^{(*)}[I(J^P)=1(0^+)]$, with the cutoff
less than 3.0 GeV, the long-range pion exchange is not sufficient
to form the bound state. However, when we add the heavier eta,
sigma, rho and omega exchanges and tune the cutoff to be 2.64 MeV,
a bound state with mass $3730.17$ MeV appears. The binding energy
relative to the $DD$ threshold is $2.64$ MeV and the corresponding
root-mean-square radius is $1.38$ fm. The channel $DD(^1S_0)$ with
a probability of $91.47\%$ dominates this state. The probability
of the channel $D^*D^*(^5D_0)$ is very small, only $0.07\%$.

Just as the $J^P=0^+$ case, with the pion-exchange potential alone
we obtain no binding solutions for the state
$D^{(*)}D^{(*)}[I(J^P)=1(1^+)]$ with the cutoff parameter less
than 3.0 GeV. When we use the OBE potential and tune the cutoff to
be $2.48$ GeV, we obtain a bound $D^{(*)}D^{(*)}[I(J^P)=1(1^+)]$
state with mass $3875.58$ MeV. The binding energy relative to the
$DD^*$ threshold is $0.27$ MeV and the corresponding
root-mean-square radius is $5.81$ fm. The dominant channel is
$[DD^*]_{+}(^3S_1)$, with a probability of $99.92\%$.

For the state $D^{(*)}D^{(*)}[I(J^P)=1(2^+)]$, it is necessary to
mention that there are five channels $DD(^1D_2)$,
$[DD^*]_{+}(^3D_2)$, $D^*D^*(^5S_2)$, $D^*D^*(^1D_2)$ and
$D^*D^*(^5D_2)$, with the quantum numbers $I(J^P)=1(2^+)$. If we
consider all the five channels, with the OBE potential we obtain a
bound state with the cutoff parameter fixed to be 2.84 GeV. The
binding energy relative to the $DD$ threshold is $12.93$ MeV.
Surprisingly, the corresponding root-mean-square radius is as
small as $0.22$ fm. The dominant channel is $D^*D^*(^5S_2)$, with
a probability of $99.5\%$. However, the probability of  the
channel $DD(^1D_2)$ is so small as $0.04\%$, which tells us that
this is not a loosely bound $DD$ state but a deeply bound
$D^*D^*$ state. With so tight a bound state, the present
meson-exchange model dose not work. Therefore, we omit the
channels $DD(^3D_2)$ and $[DD^*]_{+}(^3D_2)$ and keep the three
$D^*D^*$ channels. With the pion-exchange potential we fail to
obtain a bound state with the cutoff parameter less than 3.0 GeV.
However, when we use the OBE potential and tune the cutoff to be
$2.48$ GeV, we obtain a bound state with mass $4014.29$ MeV. The
binding energy is $2.95$ MeV and the corresponding
root-mean-square radius is $1.61$ fm. The channel $D^*D^*(^5S_2)$
with a probability of $99.93\%$ dominates this state.

The three states $D^{(*)}D^{(*)}[I(J^P)=1(0^+),1(1^+),1(2^+)]$ do not form
molecules due to the large cutoff in the present meson-exchange
model.

\renewcommand{\arraystretch}{1.0}
\begin{table*}[htp]
\centering \caption{The numerical results for the $D^{(*)}D^{(*)}$
system. ``$***$" means the corresponding
state dose not exit due to symmetry while ``$-$" means there dose
not exist binding energy with the cutoff parameter less than 3.0
GeV. The binding energies for the states
$D^{(*)}D^{(*)}[I(J^p)=0(1^+)]$ and
$D^{(*)}D^{(*)}[I(J^P)=1(1^+)]$ are relative to the threshold of
$DD^*$ while that of the state $D^{(*)}D^{(*)}[I(J^P)=1(0^+)]$ is
relative to the $DD$ threshold.} \label{numerical:DD}
\begin{tabular*}{18cm}{@{\extracolsep{\fill}}ccccccc|cccc}
\toprule[1.0pt]
        ~        &        ~        &       ~      &\multicolumn{8}{c}{$D^{(*)}D^{(*)}$}  \\
$I$              &     $J^P$       &              &\multicolumn{4}{c}{OPE}       &\multicolumn{4}{c}{OBE}  \\
\hline \multirow{10}*{0}&$0^+$            &         ~
&\multicolumn{4}{c|}{$***$}  & \multicolumn{4}{c}{$***$}\\
\cline{2-11}
      ~          &\multirow{8}*{$1^+$}&$\Lambda$(GeV) &1.05   & 1.10    &  1.15
&  1.20 &0.95    & 1.00  & 1.05  & 1.10    \\
        ~        &       ~            & B.E. (MeV)    &1.24   & 4.63    & 11.02 & 20.98 & 0.47   & 5.44  & 18.72 & 42.82  \\
        ~        &        ~        & M (MeV)        &3874.61& 3871.22 &3864.83
&3854.87&3875.38 &3870.41&3857.13&3833.03 \\
        ~        &        ~        &$r_{rms}$(fm)   &3.11   &  1.68   & 1.12
&  0.84 &4.46    &1.58   & 0.91  & 0.64   \\
        ~        &        ~        &$P_1(\%)$       &96.39  & 92.71   &88.22
& 83.34 &97.97   &92.94  & 85.64 & 77.88  \\
        ~        &        ~        &$P_2(\%)$       &0.73   & 0.72    & 0.57
& 0.42 &0.58    &0.55   &0.32   & 0.15    \\
       ~         &       ~         &$P_3(\%)$       &2.79   & 6.45    & 11.07
& 16.11&1.41    &6.42   &13.97  & 21.91    \\
       ~         &       ~         &$P_4(\%)$       &0.08   & 0.13    & 0.14
& 0.13 &0.04    &0.09   & 0.08  & 0.05    \\  \cline{2-11}
    ~            & $2^+$           &                & -     &  -      &  -
&  -   &    -    &   -  &   -   &   -     \\
\hline \multirow{23}*{1}& \multirow{7}*{$0^+$}&$\Lambda$(GeV)& -
&  -   &   -
&       -        &   2.64    & 2.66  & 2.68  & 2.70 \\
        ~        &        ~        &B.E.(MeV)      &  -     &   -  &   -
&      -         &    4.29   &12.63  & 23.30 &35.75  \\
        ~        &        ~        &M(MeV)         &  -     &   -   &  -
&       -        &   3730.17 &3721.83&3711.16&3698.71 \\
       ~         &        ~        &$r_{rms}$(fm)  &   -    &   -  &   -
&      -         &   1.38    & 0.79  & 0.58  & 0.48   \\
       ~         &        ~        & $P_1(\%)$     &   -    &   -  &   -
&      -         &   91.47   &88.26  &86.25  &84.80   \\
       ~         &        ~        & $P_2(\%)$     &   -    &   -  &   -
&      -         &   8.46    &11.64  &13.63  &  15.07 \\
       ~         &        ~        & $P_3(\%)$     &   -    &   -  &    -
&      -         &   0.07    &0.10   & 0.12  &  0.13  \\
\cline{2-11}
       ~         &\multirow{7}*{$1^+$}&$\Lambda$(GeV)&   -  &   -  &  -
&      -         &2.48    &2.50   & 2.52  &2.54     \\
        ~        &        ~        & B.E. (MeV) &   -    &    -    &   -
&      -         & 0.27   &3.82   & 9.79   &17.39    \\
       ~         &        ~        & M(MeV)     &   -    &    -    &   -
&      -         & 3875.58&3872.03&3866.06 &3858.46 \\
       ~         &        ~        & $r_{rms}$(fm)&  -   &    -    &   -
&      -         &5.81    &1.47   &0.91    &0.69     \\
       ~         &        ~        & $P_2(\%)$    &  -   &    -    &   -
&      -         &99.92   &99.90  &99.90   & 99.90   \\
        ~        &        ~        & $P_3(\%)$    &  -   &   -     &   -
&      -         &0.06    &0.06   & 0.04   & 0.04    \\
        ~        &        ~        & $P_4(\%)$    &  -   &   -     &   -
&      -         & 0.02   &0.05   & 0.06   & 0.06    \\
 \cline{2-11}
       ~         &\multirow{7}*{$2^+$}&$\Lambda$(GeV)&  - &  -    &   -
&      -         & 2.48   &2.50   & 2.52   & 2.54\\
       ~         &        ~        &B.E.(MeV)   &   -    &  -     &   -
&      -         & 2.95   & 8.86   & 16.51 & 25.54 \\
       ~         &       ~         &M(MeV)      &   -    &   -    &   -
&      -         &4014.29 &4008.38 &4000.73& 3991.70\\
       ~         &        ~        &$r_{rms}$(fm)&   -   &   -    &   -
&      -         & 1.61   & 0.92   & 0.68  & 0.56  \\
       ~         &        ~        &$P_1(\%)$   &    -   &   -    &   -
&      ~         & 99.93  & 99.94  & 99.95 & 99.95  \\
       ~         &       ~         &$P_2(\%)$   &   -   &   -    &   -
&      -         & 0.01   & 0.01   & 0.01  & 0.00  \\
       ~         &       ~         &$P_3(\%)$   &   -   &   -    &   -
&      -         & 0.06   & 0.05   & 0.04   & 0.04 \\
\bottomrule[1.0pt]
\end{tabular*}
\end{table*}

\subsubsection{$\bar{B}^{(*)}\bar{B}^{(*)}$}

With the heavy quark flavor symmetry, the potentials for the
$\bar{B}^{(*)}\bar{B}^{(*)}$ system are similar to those for the
$DD$ system. The main difference between the two systems is that
the reduced mass of the $\bar{B}^{(*)}\bar{B}^{(*)}$ system is
much larger than that of the $D^{(*)}D^{(*)}$ system. We summarize
our numerical results of the $\bar{B}^{(*)}\bar{B}^{(*)}$ system
in Table~\ref{numerical:BB}.

Similar to the charmed state $D^{(*)}D^{(*)}[I(J^P)=0(1^+)]$, the
bottomed state $\bar{B}^{(*)}\bar{B}^{(*)}[I(J^P)=0(1^+)]$ is also
very interesting. The long-range pion exchange is strong enough to
form the loosely bound $\bar{B}^{(*)}\bar{B}^{(*)}[I(J^P)=0(1^+)]$
state with the cutoff larger than $1.25$ GeV. If we tune the
cutoff to be $1.30$ GeV, the the binding energy is $1.14$ MeV and
the corresponding root-mean-square radius is $2.25$ fm. The
channel $[\bar{B}\bar{B}^*]_{-}(^3S_1)$ with a probability of
$91.83\%$ dominates this state. The probability of the D-wave is
$6.96\%$, see Table~\ref{numerical:BB}. When we use the OBE
potential, the binding becomes tighter as expected, which is
similar to its charmed partner $D^{(*)}D^{(*)}[I(J^P)=0(1^+)]$.
The numerical results suggest that the
state $\bar{B}^{(*)}\bar{B}^{(*)}[I(J^P)=0(1^+)]$ seems to be a good molecule candidate.

Different from its charmed partner,
$\bar{B}^{(*)}\bar{B}^{(*)}[I(J^P)=0(2^+)]$ can form a bound state
with the pion-exchange potential if the cutoff is tuned larger
than $2.88$ GeV. When we set the cutoff to be $2.88$ GeV, the
binding energy is $2.75$ MeV, and correspondingly, the
root-mean-square radius is $0.72$ fm. When we use the OBE
potential, we obtain the binding solutions with a smaller but more
reasonable cutoff. Unfortunately, the binding solutions depend
very sensitively on the cutoff parameter. When tune the cutoff
from $1.66$ GeV to $1.72$ GeV, the binding energy changes from
$8.30$ MeV to $73.89$ MeV. Despite the \textit{reasonable} cutoff, we can not draw a definite conclusion about the state
$\bar{B}^{(*)}\bar{B}^{(*)}[I(J^P)=0(2^+)]$ because of the strong dependence of the results on the cutoff.

The pion-exchange alone is also sufficient to form the loosely
bound $\bar{B}^{(*)}\bar{B}^{(*)}[I(J^P)=1(0^+)]$ state with the
cutoff larger than $1.70$ GeV. When we tune the cutoff from $1.70$
GeV to $1.90$ GeV, the binding energy increases from $1.05$ MeV to
$11.29$ MeV, and correspondingly, the root-mean-square radius
decreases from $2.07$ fm to $0.75$ fm. The dominant channel is
$\bar{B}\bar{B}(^1S_0)$, with a probability of
$95.35\%\sim90.29\%$. And, the probability of the channel
$D^*D^*(^5D_0)$ is $1.79\%\sim4.69\%$. Actually, two pseudoscalar
$D$-mesons can not interact with each other via exchanging a pion.
Therefore, the binding solutions totally come from the
coupled-channel effect, just as in the $\Lambda_Q\Lambda_Q$
case~\cite{Meguro:2011nr,Li:2012bt}. When we add the contributions
of the heavier eta, sigma, rho and omega exchanges, the results
change little, which implies that the eta, sigma, rho and omega
exchanges cancel with each other significantly. Although the the results depend a little sensitively on the cutoff, the state $\bar{B}^{(*)}\bar{B}^{(*)}[I(J^P)=1(0^+)]$ might also be a molecule candidate.

The state $\bar{B}^{(*)}\bar{B}^{(*)}[I(J^P)=1(1^+)]$ is also an
interesting one. When we tune the cutoff between $1.40$ GeV and
$1.70$ GeV, we also obtain a loosely bound state with the OPE
potential. The binding energy is $0.83\sim11.80$ MeV and the
corresponding root-mean-square radius is $2.36\sim0.80$ fm. The
dominant channel is $[\bar{B}\bar{B}^*]_{+}(^3S_1)$, with a
probability of $96.59\%\sim90.63\%$. Different from the isospin
singlet case, when we use the OBE potential the binding becomes
shallower. For example, with the OPE potential the binding energy
is $11.80$ MeV if the cutoff is set to be $1.70$ MeV while with
the OBE potential it is $8.29$ MeV for the same cutoff, see
Table~\ref{numerical:BB}. The present numerical results indicate that the state $\bar{B}^{(*)}\bar{B}^{(*)}[I(J^P)=1(1^+)]$ might be a good molecule candidate.

With the same reason as for the charmed case, we only keep the
three channels $\bar{B}^*\bar{B}^*(^5S_2)$,
$\bar{B}^*\bar{B}^*(^1D_2)$ and $\bar{B}^*\bar{B}^*(^5D_2)$ for
the state $\bar{B}^{(*)}\bar{B}^{(*)}[I(J^P)=0(2^+)]$. with the
pion-exchange potential, when we tune the cutoff between $1.80$
GeV and $2.10$ GeV, we obtain a loosely bound state with binding
energy $2.25\sim13.35$ MeV and root-mean-square radius
$1.48\sim0.71$ fm. The dominant channel is
$\bar{B}^*\bar{B}^*(^5S_2)$, with a probability of
$95.73\%\sim93.66\%$. When we use the OBE potential and tune the
cutoff from $1.70$ GeV to $1.76$ GeV, the binding energy changes
into $7.88\sim28.01$ MeV, and correspondingly, the
root-mean-square radius changes into $0.59\sim0.36$ fm. Similar to the $I(J^P)=1(0^+)$ case, the state $\bar{B}^{(*)}\bar{B}^{(*)}[I(J^P)=1(2^+)]$ might also be a molecule candidate.

\renewcommand{\arraystretch}{1.0}
\begin{table*}[htp]
\centering \caption{The numerical results for the
$\bar{B}^{(*)}\bar{B}^{(*)}$ system.
``$***$" means the corresponding state dose not exist due to
the symmetry. The binding energies of the states
$\bar{B}^{(*)}\bar{B}^{(*)}[I(J^P)=0(1^+)]$,
$\bar{B}^{(*)}\bar{B}^{(*)}[I(J^P)=0(2^+)]$ and
$\bar{B}^{(*)}\bar{B}^{(*)}[I(J^P)=1(1^+)]$ are relative to the
$\bar{B}\bar{B}^*$ threshold while that of the state
$\bar{B}^{(*)}\bar{B}^{(*)}[I(J^P)=1(0^+)]$ is relative to the
$\bar{B}\bar{B}$ threshold.}\label{numerical:BB}
\begin{tabular*}{18cm}{@{\extracolsep{\fill}}ccccccc|cccc}
\toprule[1.0pt]\addlinespace[3pt]
     ~           &        ~        &       ~      &\multicolumn{8}{c}{$\bar{B}^{(*)}\bar{B}^{(*)}$}\\
$I$              &     $J^P$       &
&\multicolumn{4}{c}{OPE}       &
\multicolumn{4}{c}{OBE}  \\
\hline
\multirow{15}*{0}&$0^+$            &         ~
&\multicolumn{4}{c|}{$***$}  &
\multicolumn{4}{c}{$***$}\\
 \cline{2-11}
      ~          &\multirow{8}*{$1^+$}&$\Lambda$(GeV)&1.25   & 1.30    &  1.40 &  1.50&1.10 & 1.15  & 1.20  & 1.25   \\
        ~        &       ~         & B.E. (MeV)      &0.36   & 1.14    & 4.10  &  9.44& 0.19   & 1.56  & 5.20 & 13.71\\
        ~        &        ~        & M (MeV)       &10604.08&10603.30&10600.34
&10595.00&10604.25&10602.88&10599.24&10590.73\\
        ~        &        ~        &$r_{rms}$(fm)  &3.60    &2.25    & 1.40
&  1.04 &4.86    &1.99   & 1.26  & 0.86    \\
        ~        &        ~        &$P_1(\%)$      &94.29  & 91.83   &89.03
& 83.18 &96.66   &94.17  & 92.34 &86.54    \\
        ~        &        ~        &$P_2(\%)$       &4.16   & 5.66    & 7.01
& 7.10 &2.61    &4.04   &3.54   & 1.96     \\
       ~         &       ~         &$P_3(\%)$       &0.77   &1.20     &1.74
 & 1.89  &0.36    &0.97   &3.14   &10.79    \\
       ~         &       ~         &$P_4(\%)$       &0.78   & 1.30    & 2.21
& 2.83 &0.37    &0.82   & 0.98  & 0.72    \\  \cline{2-11}
    ~            &\multirow{5}*{$2^+$}&$\Lambda$(GeV)&2.88   & 2.90    & 2.94
& 2.96  &1.66    & 1.68  & 1.70  & 1.72    \\
        ~        &        ~        &B.E.(MeV)       &2.75   & 5.18    & 10.38
& 13.15 &8.30    & 28.01 & 49.89 & 73.89   \\
        ~        &        ~        &$r_{rms}$ fm     & 0.72  & 0.68    & 0.63
& 0.61  &  0.39  & 0.35  &  0.32  & 0.31  \\
        ~        &       ~         &M(MeV)         &10601.69&10599.26&10594.06
&10591.29 &10596.14& 10576.43&10554.55&10530.55\\
        ~        &       ~         &$P_1(\%)$      &61.15   & 60.47  & 59.43
& 59.01 & 57.82  & 55.86  &  54.69  & 53.88 \\
        ~        &       ~         &$P_2(\%)$      &38.85   & 39.53  &40.57
& 40.99 &42.18   & 44.14  &  45.31  & 46.12  \\
\hline \multirow{23}*{1}& \multirow{7}*{$0^+$}&$\Lambda$(GeV)&1.70
&  1.75  & 1.80
&  1.90 &1.74&1.76  & 1.78   & 1.80    \\
        ~        &        ~        &B.E.(MeV)       &1.05    & 2.53   &  4.70
&  11.29 &2.24    &8.19   &15.67  &24.33   \\
        ~        &        ~        &M(MeV)         &10557.63&10556.15&10553.98
&10547.39      &10556.44&10550.49&10543.01&10534.35\\
       ~         &        ~        &$r_{rms}$(fm)  &2.07    & 1.40   & 1.07
& 0.75  &1.03    &0.55   & 0.42  & 0.36   \\
       ~         &        ~        & $P_1(\%)$     &95.35   & 92.82  & 90.29
& 85.46  &87.27   &83.59  &81.81  &80.66    \\
       ~         &        ~        & $P_2(\%)$     &2.02    & 3.20   & 2.42
& 6.84 &12.54   &16.26  &18.07  &  19.25   \\
       ~         &        ~        & $P_3(\%)$     &2.63    & 3.98   &  5.28
& 7.70  &0.19    &0.15   & 0.11    & 0.08    \\
\cline{2-11}
       ~         &\multirow{7}*{$1^+$}&$\Lambda$(GeV)&1.40    & 1.50   & 1.60
&  1.70 &1.66    &1.68    &1.70    & 1.72     \\
        ~        &        ~        & B.E. (MeV)     &0.83   & 1.89    & 6.46
& 11.80 & 0.55   &3.72    & 8.29   &13.92    \\
       ~         &        ~        & M(MeV)        &10603.61&10601.55&10597.98
&10592.64&10603.89&10600.72&10596.15&10590.52 \\
       ~         &        ~        & $r_{rms}$(fm) &2.36   &1.38     &1.00
&0.80   &2.26    &0.85   &0.59    &0.48     \\
       ~         &        ~        & $P_1(\%)$     &96.59  & 94.40   & 92.43
& 90.63 &99.24   &99.34  &99.51   & 99.64   \\
        ~        &        ~        & $P_2(\%)$     &1.62   & 2.35    & 2.88
& 3.29  &0.41    &0.30   & 0.20   & 0.14     \\
        ~        &        ~        & $P_3(\%)$     &1.79   & 3.25    & 4.69
& 6.08 & 0.34    &0.36   & 0.29   & 0.22     \\
 \cline{2-11}
       ~         &\multirow{7}*{$2^+$}&$\Lambda$(GeV)&1.80 & 1.90    & 2.00
& 2.10 &1.70     &1.72   & 1.74   & 1.76    \\
       ~         &        ~       &B.E.(MeV)      &2.25   & 4.74    & 8.38
& 13.35& 7.88    & 13.63 &20.36   & 28.01   \\
       ~         &       ~         &M(MeV)        &10647.95&10645.46&10641.82
&10636.85        &10642.32&10636.57&10629.84&10622.19 \\
       ~         &        ~        &$r_{rms}$     &1.48   &1.08     & 0.85
&  0.71 & 0.59   & 0.47   & 0.41   & 0.36   \\
       ~         &        ~        &$P_1(\%)$     &95.73  & 94.86   & 94.19
& 93.66 & 99.70  & 99.80  & 99.86  & 99.91  \\
       ~         &       ~         &$P_2(\%)$     &0.72   &0.87     & 0.97
& 1.06  & 0.05   & 0.03  & 0.02    & 0.01  \\
       ~         &       ~         &$P_3(\%)$     &3.55   &4.27     & 4.83
& 5.28  & 0.25   & 0.17  & 0.12    & 0.08  \\
\bottomrule[1.0pt]
\end{tabular*}
\end{table*}

\subsubsection{$D^{(*)}\bar{B}^{(*)}$}

The small binding energy and large root-mean-square radius with
reasonable cutoff parameter makes the state
$D^{(*)}\bar{B}^{(*)}[I(J^P)=0(0^+)]$ very interesting. With the
OPE potential, when we fix the cutoff between $1.08$ GeV and
$1.20$ GeV, we obtain a loosely bound state with binding energy
$0.42\sim31.35$ MeV and root-mean-square radius $4.23\sim0.55$ fm.
The dominant channel is $D\bar{B}(^1S_0)$, with a probability of
$95.32\%\sim60.03\%$. The probability of the channel
$D^*\bar{B}^*(^5D_0)$ is very small as expected, see
Table~\ref{numerical:DB}. When we add the contributions of the
heavier eta, sigma, rho and omega exchanges, the binding energy
changes by tens of MeV. It seems that the state
$D^{(*)}\bar{B}^{(*)}[I(J^P)=0(0^+)]$ might be a molecule candidate,
but not a good one because the results depend a little sensitively on the cutoff.

When the cutoff is tuned larger than $2.05$ GeV, the long-range
pion exchange is strong enough to form the loosely bound
$D^{(*)}\bar{B}^{(*)}[I(J^P)=0(1^+)]$ state. If we set the cutoff
between $2.05$ GeV and $2.20$ GeV, the binding energy relative to
the $D\bar{B}^*$ threshold is $1.21\sim 6.30$ MeV while the
root-mean-square radius is $2.75\sim1.37$ fm. The dominant channel
is $D\bar{B}^*(^3S_1)$, with a probability of
$96.81\%\sim93.31\%$. When we add the contributions of the heavier
eta, sigma, rho and omega exchanges, we obtain a loosely bound
$D^{(*)}\bar{B}^{(*)}[I(J^P)=0(1^+)]$ state with a reasonable
cutoff $1.65\sim1.80$ GeV. If we set the cutoff parameter to be
$1.70$ GeV, the binding energy is $2.83$ MeV and the
root-mean-square radius is $1.89$ fm which is comparable to the
size of the deuteron (about 2.0 fm). The channel
$D\bar{B}^*(^3S_1)$ with a probability of $95.20\%$ dominates this
state. However, the contribution of the D-wave is small, less than
$5.0\%$. With the numerical results we predict the state
$D^{(*)}\bar{B}^{(*)}[I(J^P)=0(1^+)]$ to be a molecule.

With the same reason as for the state
$\bar{B}^{(*)}\bar{B}^{(*)}[I(J^P)=0(2^+)]$, we omit the channels
$D\bar{B}(^1D_2)$, $D\bar{B}^*(^3D_2)$ and $D^*\bar{B}(^3D_2)$ for
the state $D^{(*)}\bar{B}^{(*)}[I(J^P)=0(2^+)]$. Since the
amplitudes of the channel  $D^*\bar{B}^*(^3D_2)$ scattering into
the other channels are zero, we also omit this channel in our
study. With the OPE potential, when we set the cutoff between
$2.10$ GeV and $2.30$ GeV, we obtain a bound
$D^{(*)}\bar{B}^{(*)}[I(J^P)=0(2^+)]$ state, with binding energy
$0.80\sim3.83$ fm and root-mean-square radius $3.35\sim1.73$ fm.
The channel $D^*\bar{B}^*(^5S_2)$ provides a dominant
contribution, $94.14\%\sim89.27\%$. When we use the OBE potential,
the binding energy is $0.63\sim1.79$ MeV and the root-mean-square
radius is $3.70\sim2.43$ fm for the cutoff between $1.90$ GeV and
$2.10$ GeV. Such a loosely bound state with weak dependence of the
binding solutions on the cutoff parameter is particularly
interesting. Thus the present meson-exchange approach favors
the state $D^{(*)}\bar{B}[I(J^P)=0(2^+)]$ to be a good molecule candidate.

For the state $D^{(*)}\bar{B}^{(*)}[I(J^P)=1(0^+)]$, if we tune
the cutoff parameter larger than $2.60$ GeV, the OPE potential is
sufficient to form the $D^{(*)}\bar{B}^{(*)}[I(J^P)=1(0^+)]$ bound
state. When we fix the cutoff parameter between $2.60$ GeV and
$2.90$ Gev, the binding energy is $0.14\sim17.53$ fm and the
corresponding root-mean-square radius is $7.05\sim0.75$ fm. The
dominant channel is $D\bar{B}(^1S_0)$, with a probability of
$98.68\%\sim87.27\%$. When we use the OBE potential, we obtain
binding solutions with a cutoff parameter larger than $2.22$ GeV,
see Table~\ref{numerical:DB}. The
state $D^{(*)}\bar{B}^{(*)}[I(J^P)=1(0^+)]$ might also be a molecule candidate.

For the state $D^{(*)}\bar{B}^{(*)}[I(J^P)=1(1^+)]$, when we tune
the cutoff parameter larger than $2.55$ GeV, we obtain binding
solutions with the OPE potential. If we set the cutoff parameter
to be $2.60$ GeV, the binding energy relative to the $D\bar{B}^*$
threshold is $2.43$ MeV and the corresponding root-mean-square
radius is $1.83$ fm. However, when we add the heavier eta, sigma,
rho and omega exchanges, we obtain no binding solutions with the
cutoff parameter less than $3.0$ GeV. It seems that the present
meson-exchange approach does not support the state $D^{(*)}\bar{B}^{(*)}[I(J^P)=1(1^+)]$ to be a molecule.

For the state $D^{(*)}\bar{B}^{(*)}[I(J^P)=1(2^+)]$, when we tune
the cutoff parameter as large as $2.80$ GeV, we obtain binding
solutions with the OPE potential. If we set the cutoff parameter
to be $2.90$ GeV, the binding energy is $2.00$ MeV and the
corresponding root-mean-square radius is $1.95$ fm. The dominant
channel is $D^*\bar{B}^*(^5S_2)$, with a probability of $97.56\%$.
The probability of the D-wave is $2.44\%$. When we use the OBE
potential, we obtain binding solutions with a smaller cutoff, see
Table~\ref{numerical:DB}. If we tune the cutoff to be $2.10$ GeV,
the binding energy is $0.44$ MeV. Similar to the $I(J^P)=1(0^+)$
case, the state $D^{(*)}\bar{B}^{(*)}[I(J^P)=1(2^+)]$ might also be a molecule.
\renewcommand{\arraystretch}{1.0}
\begin{table*}[htp]
\centering \caption{The numerical results for
the $D^{(*)}\bar{B}^{(*)}$ system. ``$-$" means
we obtain no binding solutions for the corresponding state with
the cutoff parameter less than 3.0 GeV. The binding energies of
the states $D^{(*)}\bar{B}^{(*)}[I(J^P)=0(0^+)]$ and
$D^{(*)}\bar{B}^{(*)}[I(J^P)=1(0^+)]$ are relative to the
$D\bar{B}^*$ threshold while those of the states
$D^{(*)}\bar{B}^{(*)}[I(J^P)=0(1^+)]$ and
$D^{(*)}\bar{B}^{(*)}[I(J^P)=1(1^+)]$ are relative to the
$D\bar{B}$ threshold.}\label{numerical:DB}
\begin{tabular*}{18cm}{@{\extracolsep{\fill}}ccccccc|cccc}
\toprule[1.0pt]\addlinespace[3pt]
       ~         &       ~         &       ~      &\multicolumn{8}{c}{$D^{(*)}\bar{B}^{(*)}$}  \\
$I$              &     $J^P$       &
&\multicolumn{4}{c}{OPE}       &
\multicolumn{4}{c}{OBE}  \\
\hline
\multirow{29}*{0}&\multirow{7}*{$0^+$}&$\Lambda$(GeV)&1.08   & 1.12    & 1.16  &  1.20&1.00 & 1.02  & 1.04  & 1.06   \\
       ~         &       ~            & B.E.(MeV)   &0.42    & 5.13    & 15.54
& 31.35& 1.92& 6.84  & 15.25 & 27.16 \\
       ~         &       ~            & M(MeV)     &7146.15 &7141.15 & 7131.03
& 7115.22&7144.65&7139.73&7131.32&7119.41\\
       ~         &        ~           & $r_{rms}$(fm)&4.23&1.27& 0.75
& 0.55 &2.07     & 1.14  & 0.78  & 0.61   \\
       ~         &        ~           & $P_1(\%)$    &95.32  & 82.45   & 69.99  & 60.03& 91.21   & 82.14 &72.93  & 64.72  \\
       ~         &        ~           & $P_2(\%)$    &4.39   &17.00    & 29.51  & 39.59& 8.47    & 17.49 & 26.75 & 35.03  \\
       ~         &         ~          & $p_3(\%)$    &0.28   &0.55     & 0.50   & 0.38 & 0.32    & 0.37  & 0.32  & 0.25   \\ \cline{2-11}
      ~          &\multirow{11}*{$1^+$}&$\Lambda$(GeV)&2.05  & 2.10    & 2.15  & 2.20 &1.65     & 1.70    &  1.75  &  1.80  \\
        ~        &       ~          & B.E. (MeV)   & 1.21   & 2.44  & 4.13  & 6.30   &0.44     & 2.83    & 10.75  & 30.86 \\
        ~        &        ~         & M (MeV)      &7191.12  &7189.89&7188.20& 7186.03&7191.89  &7189.50  &7181.58 &7161.47 \\
        ~        &        ~         &$r_{rms}$(fm)  &2.75     &2.02   & 1.63  & 1.37   &  4.37   &  1.89   &  1.05  & 0.65  \\
        ~        &        ~        &$P_1(\%)$     &96.81     &95.59   &94.42   &93.31 &98.72    & 95.20   &84.46   &65.15  \\
        ~        &        ~        &$P_2(\%)$     &0.00      &0.01    &0.01    &0.01  & 0.00    & 0.00    & 0.01  & 0.01   \\
       ~         &       ~         &$P_3(\%)$     &0.22      &0.30    &0.38   &0.46  &0.26     &1.49     &5.96   &13.96  \\
       ~         &       ~         &$P_4(\%)$     &1.10      &1.51    & 1.90  & 2.26   & 0.28    & 0.53    & 0.58  &0.34   \\
       ~         &       ~         &$P_5(\%)$     &0.29      &0.40    & 0.50  & 0.59   & 0.39    & 2.08    & 8.23  &20.08  \\
       ~         &        ~        & $P_6(\%)$    & 0.38     &0.52    & 0.67  &  0.80   &0.09     & 0.20    & 0.24  & 0.16  \\
       ~         &        ~        & $P_7(\%)$    & 1.20     &1.67    & 2.13  & 2.56   &0.26     & 0.50    & 0.52  & 0.29  \\
 \cline{2-11}
    ~            &\multirow{7}*{$2^+$}&$\Lambda$(GeV)&2.10   & 2.15   &2.20   &         2.30   &  1.90   &  1.95   &  2.00 &  2.10 \\
        ~        &        ~        &B.E.(MeV)        &0.80   & 1.31   &1.97   &         3.83   &  0.63   &  0.87   &  1.14 & 1.79 \\
        ~        &       ~         &M(MeV)        &7332.92  &7332.41  &7331.75 &7329.89& 7333.09& 7332.85 &7332.58& 7331.93\\
        ~        &       ~         &$r_{rms}$(fm) & 3.35    & 2.71    & 2.27   & 1.73  & 3.79   &  3.29   &  2.93 &  2.43  \\
        ~        &       ~         &$P_1(\%)$     &94.14    &92.83    & 91.58   & 89.27 & 96.83  &  96.41  & 96.03 & 95.36  \\
        ~        &       ~         &$P_2(\%)$     &0.25     &0.29     & 0.33   & 0.38  & 0.20   &  0.22   & 0.24  & 0.28  \\
        ~        &       ~         &$P_3(\%)$     &5.60     &6.88     & 8.09   &  10.35& 2.97   & 3.36    & 3.73  & 4.36  \\
\hline
\multirow{18}*{1}&\multirow{7}*{$0^+$}&$\Lambda$(GeV)&2.60   & 2.70    & 2.80  &  2.90  &2.22 & 2.24  & 2.26  & 2.28 \\
       ~         &       ~            & B.E.(MeV)    &0.14   & 2.56    & 8.27
& 17.53& 0.62& 6.64  & 15.45 & 26.01  \\
       ~         &       ~            & M(MeV)     &7146.43&7144 01  & 7138.30
& 7129.04&7145.95&7139.93&7131.12&7120.56\\
       ~         &        ~           & $r_{rms}$(fm)&7.05 & 1.78    & 1.04
& 0.75           &3.05   & 0.87  & 0.57  & 0.45   \\
       ~         &        ~           & $P_1(\%)$  &98.68  & 94.88   & 90.98  &87.27           & 94.14 & 87.90 &85.29  & 83.65  \\
       ~         &        ~           & $P_2(\%)$  &0.54   &2.39     & 4.36
& 6.32           & 5.83  & 12.05 & 14.65 & 16.29   \\
       ~         &         ~          & $p_3(\%)$  &0.66   &2.73     & 4.66
& 6.41           & 0.03  & 0.05  & 0.06  & 0.06   \\ [3pt]
\cline{2-11}
      ~          &\multirow{11}*{$1^+$}&$\Lambda$(GeV)& 2.55  & 2.60 & 2.65   &2.70   &   -    &    -    &    -  &    -   \\
        ~        &       ~         & B.E. (MeV)   & 0.94   & 2.43 & 4.67  & 7.67   &     -   &    -    &    -  &    -    \\
        ~        &        ~        & M (MeV)      &7191.39 &7189.90&7187.66&
7184.66 &    -   &    -    &    -  &    -     \\
        ~        &        ~        &$r_{rms}$(fm) &  2.87  & 1.83  & 1.35  &
1.08   &     -   &    -    &    -  &    -   \\
        ~        &        ~        &$P_1(\%)$     & 96.13  &93.82  &91.53  & 89.31  &     -   &    -    &    -  &    -   \\
        ~        &        ~        &$P_2(\%)$     &0.00    &0.01   &0.01    &
0.01   &     -   &    -    &    -  &    -   \\
       ~         &       ~         &$P_3(\%)$     &0.99    &1.64   &2.33    &
3.03   &     -   &    -    &    -  &    -   \\
       ~         &       ~         &$P_4(\%)$     &0.70    &1.09   &1.45   & 1.77   &     -   &     -   &    -  &     -   \\
       ~         &       ~         &$P_5(\%)$     &0.92    &1.48   &2.03   & 2.57   &    -    &     -   &    -  &     -   \\
       ~         &        ~        & $P_6(\%)$    & 0.26   & 0.41  & 0.54  & 0.66   &    -    &     -   &    -  &     -   \\
       ~         &        ~        & $P_7(\%)$    & 0.99   & 1.56  & 2.12  & 2.66   &    -    &     -   &    -  &     -   \\
\cline{2-11}
       ~       &\multirow{7}*{$2^+$}&$\Lambda$(GeV)& 2.80   & 2.85  & 2.90  &
2.95   & 2.10    &  2.12   & 2.14  & 2.16   \\
       ~         &         ~       & B.E.(MeV)    & 0.49   & 1.11  & 2.00  &
3.17   & 0.44    &  4.64   & 10.86 & 18.45  \\
       ~         &         ~       & M(MeV)       & 7333.23& 7332.61& 7331.72&
7330.55&7333.28  &7329.08  &7322.86&7315.27 \\
       ~         &         ~       &$r_{rms}$(fm) & 3.80   & 2.57   & 1.95  &
1.58   &  3.51   & 1.03    & 0.68  & 0.54   \\
      ~          &         ~       &$P_1(\%)$     & 98.57  & 98.03  & 97.56 &
97.15  &  99.86  & 99.89   & 99.93 & 99.95  \\
       ~         &         ~       &$P_2(\%)$     &0.24    & 0.33   & 0.40  &
0.47   &  0.02   &  0.02   &  0.01 &  0.01  \\
       ~         &         ~       &$P_3(\%)$     & 1.20   & 1.65   & 2.04  &
2.38   &  0.12   &  0.09   &  0.06 &  0.04  \\
\bottomrule[1.0pt]
\end{tabular*}
\end{table*}

\subsection{The Results for The Systems with Strangeness $S=1$}

For the systems with strangeness $S=1$, there dose not exist the
long-range pion exchange, but there are heavier eta, sigma, $K$
and $K^*$ exchanges. We summarize our numerical results in
Table~\ref{numerical:DDBBs} for $(DD)_s$ and $(\bar{B}\bar{B})_s$
and in Table~\ref{numerical:DBs} for $(DB)_s$.

\subsubsection{$(D^{(*)}D^{(*)})_s$ and $(\bar{B}^{(*)}\bar{B}^{(*)})_s$}

When we fix the cutoff parameter between $2.70$ GeV and $2.76$ GeV,
we obtain a bound state of $(D^{(*)}D^{(*)})_s[J^P=0^+]$ with mass
between $3832.06$ MeV and $3802.15$ MeV and a large root-mean-square
radius $1.53\sim0.50$ fm. The dominant channel is $DD_s(^1S_0)$ with
a probability of $92.85\%\sim85.82\%$ and the second largest channel
$D^*D^*(^1S_0)$ contributes $7.10\%\sim14.07\%$. The contribution of
the channel $D^*D^*_s(^5D_0)$ is very small, only $0.05\%\sim
0.11\%$. Now the cutoff parameter is $2.70$ GeV which is twice as
large as that for the deuteron ($0.8\sim1.5$ GeV~\cite{Machleidt}).
Besides, the binding solutions depend sensitively on the cutoff. It
seems that the present meson-exchange approach dose not support the
state $(D^{(*)}D^{(*)})_s[J^p=0^+]$ to be a molecule candidate.

In the corresponding bottom sector, we also obtain a bound state
of $(\bar{B}^{(*)}\bar{B}^{(*)})_s[J^P=0^+]$ with a smaller and
more reasonable cutoff, $1.82\sim1.88$ GeV. This can be easily
understood since the mass of the bottomed mesons are much
larger than those of the charmed mesons and the effective
potentials for the two systems are similar. When we tune the cutoff
to be $1.82$ GeV, the binding energy is $0.56$ GeV and the
root-mean-square is $2.28$ fm which is comparable to that of the
deuteron (about $2.0$ fm), see Table~\ref{numerical:DDBBs}. The
numerical results indicate that the state $(\bar{B}^{(*)}\bar{B}^{(*)})_s[J^P=0^+]$
might be a molecule candidate although not a good one.

For the charmed state $(D^{(*)}D^{(*)})_s[J^P=1^+]$, when we set the cutoff to be $1.44$ GeV, the binding
energy is $5.43$ MeV and the root-mean-square radius is $1.36$ fm.
The closeness of the thresholds
for $DD_s^*$ and $D^*D_s$ makes the channels $DD_s^*(^3S_1)$ and
$D^*D_s(^3S_1)$ provide the comparable and main contributions,
$45.80\%$ for $DD_s^*(^3S_1)$ and $51.31\%$ for $D^*D_s(^3S_1)$.
The D-wave channel $D^*D^*_s(^3D_1)$ and $D^*D^*_s$ provide almost
vanishing contributions, $0.02\%$ for $D^*D^*_s(3D_1)$ and
$0.00\%$ for $D^*D^*_s(^5D_1)$. This is because of the large mass
gap between $D^*D^*_s$ and $D^*D_s$ and the strong repulsive
interaction coming from the centrifugal potential of the D-wave.
The numerical results suggest that the state $(D^{(*)}D^{(*)})_s[J^P=1^+]$ might also be a molecule candidate. The results of the state
$(\bar{B}^{(*)}\bar{B}^{(*)})_s[J^P=1^+]$) are similar to those of the state $(D^{(*)}D^{(*)})_s[J^P=1^+]$, but with smaller cutoff $1.10\sim1.16$ GeV and weaker dependence on the cutoff, see Table~\ref{numerical:DDBBs}.  Thus the state $(\bar{B}^{(*)}\bar{B}^{(*)})_s[J^P=1^+]$) seems to be a good molecule candidate

For the state $(D^{(*)}D^{(*)})_s[J^P=2^+]$, we obtain a bound
state with binding energy $1.54\sim22.29$ MeV when we tune the
cutoff to be $2.54\sim2.60$ GeV. The root-mean square radius is
$2.28\sim 0.60$ fm. The dominant channel is $D^*D^*_s(^5S_2)$,
with a probability of $99.98\%$. However, the contributions of the
D-wave are negligible. The numerical results suggest that the present meson-exchange approach does not support $(D^{(*)}D^{(*)})_s[J^P=2^+]$ to be a molecule due to the large cutoff.

For the state $(\bar{B}^{(*)}\bar{B}^{(*)})_s[J^P=2^+]$, if we fix
the cutoff parameter between $1.76$ GeV and $1.82$ GeV, the
binding energy is $0.92\sim16.34$ MeV and the root-mean-square
radius is $1.70\sim0.44$ fm. The dominant channel is
$\bar{B}^*\bar{B}^*_s(^5S_2)$ with a probability of
$99.76\%\sim99.85\%$. Such a loosely bound state is very interesting. And it might be a good molecule candidate.

\renewcommand{\arraystretch}{1.0}
\begin{table*}[htp]
\centering \caption{The numerical results for the
$(D^{(*)}D^{(*)})_s/(\bar{B}^{(*)}\bar{B}^{(*)})_s$ systems. The binding energy of the state
$(D^{(*)}D^{(*)})_s/(\bar{B}^{(*)}\bar{B}^{(*)})_s[J^P=0^+]$ is
relative to the $DD_s$ threshold while that of the state
$(D^{(*)}D^{(*)})_s/(\bar{B}^{(*)}\bar{B}^{(*)})_s[J^P=1^+]$ is
relative to the $D^*D_s/\bar{B}^*\bar{B}_s$ threshold.}
\label{numerical:DDBBs}
\begin{tabular*}{18cm}{@{\extracolsep{\fill}}cccccc|cccc}
\toprule[1.0pt]\addlinespace[3pt]
    $J^P$           &              & \multicolumn{4}{c|}{$(D^{(*)}D^{(*)})_s$}    &
\multicolumn{4}{c}{$(\bar{B}^{(*)}\bar{B}^{(*)})_s$}\\
\hline \multirow{7}*{$0^+$}&$\Lambda$(GeV)&   2.70   & 2.72  &
2.74  & 2.76  &
1.82   & 1.84       &   1.86  & 1.88 \\
     ~              & B.E.(MeV)    &  3.66    & 11.31 & 21.47 & 33.57 &
0.56   & 5.27       & 12.08   & 20.32 \\
       ~            & M(MeV)       &3832.06   &3824.41&3814.25&3802.15&
10645.08&10640.37   &10633.56 &10625.32\\
        ~           & $r_{rms}$(fm)&1.53      & 0.85  & 0.62  & 0.50  &
2.28   & 0.71       & 0.48    & 0.39   \\
        ~           & $P_1(\%)$    & 92.85    & 89.58 &87.42  & 85.82 &
92.78  & 86.34      & 83.69   &82.06  \\
        ~           & $P_2(\%)$    & 7.10     & 10.34 &12.49  & 14.07 &
7.12   & 13.54      & 16.21   & 17.86 \\
         ~          & $p_3(\%)$    & 0.05     & 0.08  & 0.09  & 0.11  &
0.10   & 0.12       & 0.10    & 0.07   \\
\hline \multirow{11}*{$1^+$}&$\Lambda$(GeV)&1.44     & 1.46  &
1.48  & 1.50  &
1.10   & 1.12    &  1.14  &  1.16 \\
          ~         & B.E. (MeV)   & 5.43     & 10.19 & 16.31 & 23.84 &
0.67   & 3.21    & 7.17   & 12.52 \\
        ~        & M (MeV)         &3971.68   &3966.92&3960.80&3953.27&
10690.73&10688.19&10684.23&10678.88\\
           ~        &$r_{rms}$(fm) &1.36      &1.05   & 0.87  & 0.74  &
2.19    &1.10     & 0.81   & 0.66  \\
        ~        &$P_1(\%)$        &45.80     &47.33  &47.81  &47.85  &
24.53  &36.91    &41.11   &42.75  \\
        ~        &$P_2(\%)$        &0.14      &0.15   &0.15   &0.14   &
0.22   & 0.32    & 0.34   & 0.33   \\
       ~         &$P_3(\%)$        &51.31     &48.83  &47.40  &46.41  &
72.97  &58.66    &52.92   &49.87 \\
       ~         &$P_4(\%)$        &0.12      &0.13   & 0.13  & 0.13  &
0.36  & 0.42    & 0.40   & 0.36  \\
       ~         &$P_5(\%)$        &2.61      &3.54   & 4.49  & 5.44  &
1.83   & 3.55    & 5.09   & 6.54  \\
        ~        & $P_6(\%)$       & 0.02     & 0.02  & 0.02  & 0.02  &
0.09   & 0.13    & 0.14  & 0.15   \\
        ~        & $P_7(\%)$      & 0.00   & 0.00  & 0.00  & 0.00    &
0.01   & 0.00    & 0.00  & 0.00    \\
\hline \multirow{7}*{$2^+$}&$\Lambda$(GeV)&2.54    &2.56   & 2.58
& 2.60   &
1.76   & 1.78    & 1.80  & 1.82    \\
       ~         &B.E.(MeV)        & 1.54   & 6.59  & 13.67& 22.29   &
0.92   &  4.74   & 9.98  & 16.34   \\
        ~        & M(MeV)          &4119.38 &4114.33&4107.25&4098.63 &
10739.60 &10735.82&10730.52&10724.26\\
        ~        & $r_{rms}$(fm)   &2.28    & 1.07  & 0.75  & 0.60   &
1.70    & 0.75   & 0.54   &  0.44   \\
        ~        & $P_1(\%)$       &99.98   & 99.97 & 99.97 & 99.97  &
99.76   &  99.79 & 99.85  & 99.85   \\
        ~        & $P_2(\%)$       &0.00    & 0.00  & 0.00  & 0.00   &
0.04    &  0.03  & 0.02   & 0.02   \\
        ~        & $P_3(\%)$       &0.02    & 0.03  & 0.03  & 0.03   &
0.20    &  0.17  & 0.13   & 0.10   \\
\bottomrule[1.0pt]
\end{tabular*}
\end{table*}

\subsubsection{ $(D^{(*)}\bar{B}^{(*)})_s$}

For the state $(D^{(*)}\bar{B}^{(*)})_s[J^P=0^+]$, if we set the
cutoff parameter to be $1.28$ GeV, the binding energy is $6.72$
MeV and correspondingly, the root-mean-square radius is $0.92$ fm.
The probability of the channel $D\bar{B}_s(^1S_0)$ is $50.10\%$
while that of the channel $D_s\bar{B}(^1S_0)$ is $25.66\%$.
However, if the binding energy is $68.73$ MeV, these two channels
provide comparable contributions, $25.30\%$ for
$D\bar{B}_s(^1S_0)$ and $23.02\%$ for $D_s\bar{B}(^1S_0)$. Given
that the mass gap between $D\bar{B}_s$ and $D_s\bar{B}$ is around
$14.5$ MeV, for the binding energy comparable to this value, the
mass gap plays an important role in the formation of the loosely
bound state. However, if the binding energy is as large as $68.73$
MeV, which is much larger than the mass gap, the important effect
of the mass gap is gone, which is similar to the case of
$X$(3872)~\cite{Li:2012cs}. Despite \textit{reasonable} cutoff, we can not draw a definite conclusion about the state $(D^{(*)}\bar{B}^{(*)})_s[J^P=0^+]$ due to the strong dependence of the results on the cutoff.

It is necessary to mention that there exist fourteen channels for the state $(D^{(*)}\bar{B}^{(*)})_s[J^P=1^+]$. It is very hard
to solve a $14\times14$ matrix Shr\"odinger equation. Due to the
large mass gap between the threshold of $D^*\bar{B}^*_s$(or
$D^*_s\bar{B}^*)$ and that of $D\bar{B}_s^*$ and the strong
repulsive interaction coming from the D-wave, we expect the
channels $D^*\bar{B}_s^*(^3D_1)$, $D^*\bar{B}_s^*(^5D_1)$,
$D^*_s\bar{B}^*(^3D_1)$ and $D^*_s\bar{B}^*(^5D_1)$ to provide
negligible contributions, which can be clearly seen from the
previous calculation. Therefore, we omit these four channels in
our calculation. We obtain a bound state of $(D^{(*)}\bar{B}^{(*)})_s[J^P=1^+]$ with the cutoff larger than 1.24 GeV and the results are similar to those of $(D^{(*)}\bar{B}^{(*)})_s[J^P=0^+]$, see Table~\ref{numerical:DBs}.

For the state $(D^{(*)}\bar{B}^{(*)})_s[J^P=2^+]$, if the cutoff
parameter is fixed to be $2.20$ GeV, the binding energy relative
to threshold of $D^*\bar{B}^*_s$ is $1.23$ MeV and the
root-mean-square radius is $2.25$ fm. The dominant channel is
$D^*\bar{B}^*_s(^5S_2)$, with a probability of $88.81\%$. The
other S-wave channel $D^*_s\bar{B}^*(^5S_2)$ provides the second
largest contribution, $10.90\%$, and the total contribution of the
D-wave is $0.3\%$. It seems that the present meson-exchange approach supports $(D^{(*)}\bar{B}^{(*)})_s[J^P=2^+]$ to be a molecule candidate, but not a good one.

\begin{table*}[htp]
\centering \caption{The numerical results for the
$(D^{(*)}\bar{B}^{(*)})_s$ system. The binding
energy of the state $(D^{(*)}\bar{B}^{(*)})_s[J^P=0^+]$ is
relative to the $D\bar{B}_s$ threshold while those of the states
$(D^{(*)}\bar{B}^{(*)})_s[J^P=1^+]$ and
$(D^{(*)}\bar{B}^{(*)})_s[J^P=2^+]$ correspond to the thresholds
of $D\bar{B}_s^*$ and $D^*\bar{B}_s^*$ respectively.}
\label{numerical:DBs}
\begin{tabular*}{18cm}{@{\extracolsep{\fill}}cccccc}
\toprule[1.0pt]\addlinespace[3pt]
    $J^P$           &              & \multicolumn{4}{c}{$(D^{(*)}\bar{B}^{(*)})_s$} \\
\hline
\multirow{10}*{$0^+$}&$\Lambda$(GeV)&   1.28   & 1.30  & 1.32  & 1.34 \\
     ~              & B.E.(MeV)    &   6.72   & 22.10 & 43.11 & 68.73\\
       ~            & M(MeV)       &7226.81   &7211.43&7190.42&7164.80\\
        ~           & $r_{rms}$(fm)&0.92      & 0.55  & 0.43  & 0.36 \\
        ~           & $P_1(\%)$    &50.10     & 36.04 &29.33  & 25.30\\
        ~           & $P_2(\%)$    &25.66     & 26.98 &25.07  & 23.02\\
         ~          & $p_3(\%)$    &12.03     & 18.55 &22.96  & 26.05\\
      ~             & $P_4(\%)$    &12.06     & 18.30 & 22.56 & 25.58\\
         ~          & $P_5(\%)$    & 0.07     & 0.05  & 0.04  & 0.03 \\
         ~          & $P_6(\%)$    & 0.08     & 0.06  & 0.04  & 0.03 \\
\hline
\multirow{14}*{$1^+$}&$\Lambda$(GeV)& 1.24    & 1.26  &  1.28 & 1.30 \\
         ~          & B.E.(MeV)     & 2.50    & 14.97 & 32.88 & 54.86\\
         ~          & M(MeV)        & 7280.13 &7267.66&7249.75&7227.77\\
         ~          &$r_{rms}$(fm)  & 1.45    & 0.63  & 0.47  & 0.39  \\
         ~          &$P_1(\%)$      & 19.11   & 23.61 & 22.14 & 20.40 \\
         ~          &$P_2(\%)$      & 58.10   & 35.98 & 28.14 & 23.97 \\
         ~          &$P_3(\%)$      & 4.65    & 8.04  &  9.71 & 10.69 \\
         ~          &$P_4(\%)$      & 4.63    & 8.16  & 9.86  & 10.80 \\
         ~          &$P_5(\%)$      & 6.73    & 11.95 & 14.87 & 16.84 \\
         ~          &$P_6(\%)$      & 6.72    & 12.21 & 15.26 & 17.28 \\
         ~          &$P_7(\%)$      & 0.00    & 0.00  & 0.00  & 0.00  \\
         ~          &$P_8(\%)$      & 0.00    & 0.00  & 0.00  & 0.00  \\
         ~          &$P_9(\%)$      & 0.04    & 0.03  & 0.02  & 0.01  \\
         ~          &$P_{10}(\%)$   & 0.03    & 0.02  & 0.02  & 0.01  \\
\hline
\multirow{10}*{$2^+$}&$\Lambda$(GeV)&2.20     & 2.22  & 2.24  & 2.26 \\
         ~          & B.E.(MeV)    & 1.23     & 5.68  & 12.65 & 21.29\\
         ~          & M(MeV)       & 7422.79  &7418.34&7411.37&7402.73\\
         ~          & $r_{rms}$(fm)& 2.25     & 0.95  & 0.62  & 0.49  \\
         ~          & $P_1(\%)$    & 10.90    & 26.64 & 36.08 & 40.94 \\
         ~          & $P_2(\%)$    & 88.81    & 73.18 & 63.83 & 59.01 \\
         ~          & $P_3(\%)$    & 0.02     & 0.01  & 0.01  & 0.00  \\
         ~          & $P_4(\%)$    & 0.01     & 0.00  & 0.00  & 0.00  \\
         ~          & $P_5(\%)$    & 0.18     & 0.12  & 0.06  & 0.03  \\
         ~          & $P_6(\%)$    & 0.09     & 0.05  & 0.02  & 0.01  \\
\bottomrule[1.0pt]
\end{tabular*}
\end{table*}

\subsection{The Results for The Systems with Strangeness $S=2$}

For the systems with strangeness $S=2$, there dose not exist the
long-range pion exchange, but there are mediate-range sigma and
eta exchanges and the short-range phi exchange. We summarize the
numerical results for the systems $(D^{(*)}D^{(*)})_{ss}$ and
$(\bar{B}^{(*)}\bar{B}^{(*)})_{ss}$ in
Table~\ref{numerical:DDBBss} and for the system
$(D^{(*)}\bar{B}^{(*)})_{ss}$ in Table~\ref{numerical:DBss}.

\subsubsection{$(D^{(*)}D^{(*)})_{ss}$ and $(\bar{B}^{(*)}\bar{B}^{(*)})_{ss}$}

If we fix the cutoff parameter between $2.76$ GeV and $2.82$ GeV,
we obtain a bound state of $(D^{(*)}D^{(*)})_{ss}[J^P=0^+]$ with
binding energy $2.43\sim28.53$ MeV and root-mean-square radius
$1.92\sim0.55$ fm. The dominant channel is $D_sD_s(^1S_0)$, with a
probability of $94.69\%\sim87.27$. The contribution of the D-wave
channel $D^*_sD^*_s(^5D_0)$ is very small as expected, less than
$0.1\%$. In the bottomed sector, we obtain binding solutions with
a smaller but more reasonable cutoff $1.90\sim1.96$ GeV. As one
can easily read off from Table~\ref{numerical:DDBBss}, when we set
the cutoff to be $1.90$ GeV, the binding energy is $2.27$ MeV and
the root-mean-square is $1.17$ fm. The dominant channel is
$\bar{B}_s\bar{B}_s(^1S_0)$, with a probability of $90.83\%$.
However, the D-wave channel $\bar{B}^*\bar{B}^*(^5D_0)$ provides a
negligible contribution, $0.29\%$. Based on the numerical results, the state $\bar{B}_s\bar{B}_s(^1S_0)$ might be a molecule whereas the state $(D^{(*)}D^{(*)})_{ss}[J^P=0^+]$ might not be.

For the state $(D^{(*)}D^{(*)})_{ss}[J^P=1^+]$, when we tune the
cutoff parameter between $2.62$ GeV and $2.68$ GeV, we obtain
binding energy $0.41\sim16.57$ MeV and root-mean-square radius
$4.64\sim0.71$ fm. The dominant channel is
$[D_sD_s^*]_{+}(^3S_1)$, with a probability of
$99.98\%\sim99.96\%$. In the corresponding bottomed case, we
obtain a loosely bound state of $(\bar{B}\bar{B})_{ss}[J^P=1^+]$
with binding energy $0.83\sim11.98$ MeV and root-mean-square
radius $1.95\sim0.54$ fm when we set the cutoff parameter between
$1.82$ GeV and $1.88$ GeV. Similar to the charmed case, the
dominant channel is $[\bar{B}_s\bar{B}^*_s]_{+}(^3S_1)$, providing
a contribution of $99.41\%\sim99.29\%$. Similar to the $J^P=0^+$ case, it seems that the state $(B^{(*)}B^{(*)})_{ss}[J^P=1^+]$ might be a molecule whereas the state $(D^{(*)}D^{(*)})_{ss}[J^P=1^+]$ might not be.

For the $J^P=2^+$ case, the results are very similar to
those of the $J^P=1^+$ case, see~\ref{numerical:DDBBss}.

\begin{table*}[htp]
\centering \caption{The numerical results for the
$(D^{(*)}D^{(*)})_{ss}/(\bar{B}^{(*)}\bar{B}^{(*)})_{ss}$ systems. The binding energy of the state
$(D^{(*)}D^{(*)})_{ss}/(\bar{B}^{(*)}\bar{B}^{(*)})_{ss}[J^P=0^+]$
corresponds to the $D_sD_s/\bar{B}_s\bar{B}_s$ threshold while
that of the state
$(D^{(*)}D^{(*)})_{ss}/(\bar{B}^{(*)}\bar{B}^{(*)})_{ss}$
corresponds to the threshold of $D_sD_s^*/\bar{B}_s\bar{B}_s^*$.}
\label{numerical:DDBBss}
\begin{tabular*}{17cm}{@{\extracolsep{\fill}}cccccc|cccc}
\toprule[1.0pt]\addlinespace[3pt]
     $J^P$          &              & \multicolumn{4}{c|}{$(D^{(*)}D^{(*)})_{ss}$}    &
\multicolumn{4}{c}{$(\bar{B}^{(*)}\bar{B}^{(*)})_{ss}$}\\
\hline \multirow{7}*{$0^+$}&$\Lambda$(GeV)&2.76 & 2.78  & 2.80&
2.82  &1.90   & 1.92    &   1.94  & 1.96 \\
      ~            & B.E.(MeV)    &2.43   & 8.68  & 17.56 & 28.53 &2.27   & 6.97    &  13.30   & 20.99 \\
       ~           & M(MeV)       &3934.55&3928.30&3919.42&
3908.45&10730.33&10725.63&10719.30&10711.61\\
       ~           & $r_{rms}$(fm)&1.92   & 1.00  & 0.70  & 0.55  &1.17   & 0.67    & 0.49     & 0.40   \\
       ~           & $P_1(\%)$   & 94.69  & 91.37  &89.04  & 87.27 &90.83  & 87.05   & 84.67    &83.00  \\
       ~           & $P_2(\%)$   & 5.29   & 8.59   & 10.91 & 12.68 &8.88  & 12.65    & 15.05   & 16.76 \\
        ~          & $p_3(\%)$   & 0.02   & 0.03   & 0.04  & 0.05   &0.29     & 0.30   & 0.28      & 0.24   \\
\hline
\multirow{7}*{$1^+$}&$\Lambda$(GeV)&2.62    & 2.64   & 2.66  & 2.68   &1.82     & 1.84   &  1.86   &  1.88 \\
          ~         & B.E. (MeV)   & 0.41   & 3.75   & 9.32  & 16.57  &0.83     & 3.39   & 7.16    & 11.98 \\
          ~        & M (MeV)        &4080.38 &4077.04&4071.47&
4064.22&10778.27  &10775.71&10771.94&10767.12\\
         ~        &$r_{rms}$(fm)   & 4.64   &1.49   & 0.94 & 0.71   &1.95     &0.96    & 0.68   & 0.54  \\
         ~        &$P_1(\%)$       &99.98   &99.96  &99.96&
99.96  &99.41    &99.29   &99.33   &99.41  \\
        ~        &$P_2(\%)$       &0.01    &0.02   &0.02 &
0.01   &0.24     & 0.27   & 0.24   & 0.20   \\
        ~        &$P_3(\%)$       & 0.01   & 0.02  &0.02 &
0.03   &0.35     &0.44    &0.43    &0.39 \\
\hline
\multirow{7}*{$2^+$}&$\Lambda$(GeV)&2.60   &2.62  & 2.64  & 2.66 &1.82   & 1.84   & 1.86  & 1.88    \\
       ~         &B.E.(MeV)     & 0.86  & 4.79 &10.81 &18.46 &0.22   &  2.48  & 6.24  &11.15   \\
        ~        & M(MeV)   &4223.74&4219.81&4213.79&4206.14
&10830.61&10828.32&10824.62&10819.72\\
        ~        & $r_{rms}$(fm)&3.13 & 1.29  & 0.86  & 0.66 &3.86  & 1.10    & 0.71   &  0.55   \\
        ~        & $P_1(\%)$    &99.98   & 99.97 & 99.98 & 99.98  &99.75    &  99.61 & 99.64    & 99.70   \\
        ~        & $P_2(\%)$    &0.00    & 0.00  & 0.00  & 0.00   &0.04     &  0.06  & 0.05     & 0.04   \\
        ~        & $P_3(\%)$    &0.02    & 0.02  & 0.02  & 0.02   &0.21     &  0.33  & 0.30     & 0.26   \\
\bottomrule[1.0pt]
\end{tabular*}
\end{table*}

\subsubsection{$(D^{(*)}\bar{B}^{(*)})_{ss}$}

For the state $(D^{(*)}\bar{B}^{(*)})_{ss}[J^P=0^+]$, we obtain
binding energy $0.64\sim21.72$ MeV and root-mean-square radius
$3.11\sim0.52$ fm with the cutoff parameter fixed between
$2.36$ GeV and $2.42$ GeV. The channel $D_s\bar{B}_s(^1S_0)$ with
a probability of $96.05\%\sim 86.20\%$ is the dominant channel.
The probability of the channel $D_s^*\bar{B}_s^*(^5D_0)$ is very
small, about $0.03\%$, see Table~\ref{numerical:DBss}.

For the state $(D^{(*)}\bar{B}^{(*)})_{ss}[J^P=1^+]$, when we fix
the cutoff parameter between $2.34$ GeV and $2.40$ GeV we obtain
binding energy $1.47\sim 23.41$ MeV and corresponding
root-mean-square radius $2.02\sim0.50$ fm. The dominant channel is
$D_s\bar{B}_s^*(^3S_1)$, with a probability of
$93.37\%\sim83.13\%$. However, the total contribution of the
D-wave is very small, less than $0.1\%$, see
Table~\ref{numerical:DBss}.

Very similar to the $J^P=0^+$ case, we obtain a bound state of
$(D^{(*)}\bar{B}^{(*)})_{ss}[J^P=2^+]$ with binding energy
$2.98\sim22.29$ MeV and root-mean-square radius $1.35\sim0.51$ fm.
The dominant channel is $D_s^*\bar{B}_s^*(^5S_2)$, with a
probability of $99.98\%$. The numerical results indicate that the present meson-exchange approach seems to support all of the three states to be molecule candidates, but not good ones since the results depend a little sensitively on the cutoff.

\begin{table*}[htp]
\centering \caption{The numerical results for the
$(D^{(*)}\bar{B}^{(*)})_{ss}$ system. The binding
energy of the state $(D^{(*)}\bar{B}^{(*)})_{ss}[J^P=0^+]$ is
relative to the threshold of $D_s\bar{B}_s$ while that of the
state $(D^{(*)}\bar{B}^{(*)})_{ss}[J^P=1^+]$ corresponds to the
$D_s\bar{B}_s^*$ threshold.}\label{numerical:DBss}
\begin{tabular*}{18cm}{@{\extracolsep{\fill}}cccccc}
\toprule[1.0pt]\addlinespace[3pt]
     $J^P$          &              & \multicolumn{4}{c}{$(D^{(*)}\bar{B}^{(*)})_{ss}$}    \\
\hline
\multirow{7}*{$0^+$}&$\Lambda$(GeV)&2.36   & 2.38  & 2.40  &  2.42 \\
       ~            & B.E.(MeV)    &0.64   & 5.16  & 12.43 & 21.72 \\
      ~             & M(MeV)       &7334.15&7329.63&7322.36&7313.07\\
       ~            & $r_{rms}$(fm)&3.11   & 1.06  & 0.68  & 0.52  \\
       ~            & $P_1(\%)$    &96.05  & 91.15 &88.25  & 86.20 \\
        ~           & $P_2(\%)$    & 3.94  & 8.82  & 11.72 & 13.17 \\
        ~           & $p_3(\%)$    & 0.02  & 0.03  & 0.03  & 0.03  \\
\hline
\multirow{11}*{$1^+$}&$\Lambda$(GeV)&2.34   & 2.36  & 2.38  & 2.40  \\
        ~            & B.E. (MeV)   &1.47   & 6.63  & 14.13 & 23.41 \\
            ~        & M (MeV)      &7382.42&7377.26&7369.76&7360.48\\
            ~        &$r_{rms}$(fm) & 2.02  &0.92   & 0.63  & 0.50  \\
        ~            &$P_1(\%)$     &93.37  &88.46  &85.36  &83.13  \\
           ~         &$P_2(\%)$     &0.00   &0.00   &0.00   & 0.00  \\
          ~          &$P_3(\%)$     &3.85   &6.95   &9.07   &10.69  \\
          ~          &$P_4(\%)$     &0.01   &0.01   & 0.01  & 0.01  \\
          ~          &$P_5(\%)$     &2.75   &4.55   & 5.54  & 6.14  \\
          ~          &$P_6(\%)$     &0.00   &0.00   & 0.00  & 0.00  \\
          ~          &$P_7(\%)$     &0.02   &0.02   & 0.02  & 0.02  \\
\hline
\multirow{7}*{$2^+$} &$\Lambda$(GeV)&2.26   &2.28   & 2.30  & 2.32  \\
         ~           &B.E.(MeV)     &2.98   &7.96   &14.47  &22.29  \\
            ~        & M(MeV)       &7524.72&7519.74&7513.23&7505.41\\
            ~        & $r_{rms}$(fm)&1.35   &0.82   &0.62   &0.51   \\
         ~           & $P_1(\%)$    &99.93  &99.94  &99.95  &99.96  \\
            ~        & $P_2(\%)$    &0.01   & 0.01  & 0.01  & 0.01  \\
            ~        & $P_3(\%)$    &0.06   & 0.05  & 0.04  & 0.03  \\
\bottomrule[1.0pt]
\end{tabular*}
\end{table*}

\section{conclusion}\label{conclusion}

In the present paper, we investigate the possible molecular states
composed of two heavy flavor mesons, including $D^{(*)}D^{(*)}$,
$\bar{B}^{(*)}\bar{B}^{(*)}$ and $D^{(*)}\bar{B}^{(*)}$ with
strangeness $S=0$, $1$ and $2$. In our study, we take into account
the S-D mixing which plays an important role in the formation of
the loosely bound deuteron, and particularly, the coupled-channel
effect in the flavor space.

In order to make clear the role of the long-range pion exchange in
the formation of the loosely bound states, we give the numerical
results with the one-pion-exchange potential for the system with
strangeness $S=0$, as well as the numerical results with the
one-boson-exchange potential.

In our study, we notice that for some systems, such
as $D^{(*)}D^{(*)}[I(J^P)=0(1^+)]$, the probability of the D-wave is
very small. What's more, the contributions of the D-wave channel
with larger threshold is almost negligible for the system with a
large mass gap among the thresholds of different channels.
We also notice that when the binding energy is comparable to or
even smaller than the mass gap, the effect of the mass gap will be
magnified by the small binding energy, which is similar to the
$X(3872)$ case~\cite{Li:2012cs}.

In the sector with strangeness $S=0$, our results favor that
$D^{(*)}D^{(*)}[I(J^P)=0(1^+)]$,
$\bar{B}^{(*)}\bar{B}^{(*)}[I(J^P)=0(1^+),1(1^+)]$
and $D^{(*)}\bar{B}^{(*)}[I(J^P)=0(1^+),0(2^+)]$ are good molecule
candidates. For these states, the long-range pion exchange is
strong enough to form the loosely bound states, and the
medium-range eta and sigma exchanges and the short-range rho and
omega exchanges are helpful to strengthen the binding.
The states $\bar{B}^{(*)}\bar{B}^{(*)}[I(J^P)=1(0^+),1(2^+)]$ and
$D^{(*)}\bar{B}^{(*)}[I(J^P)=0(0^+),1(0^+),1(2^+)]$ might also be
molecules although the results depend a little sensitively on the cutoff.
However, it seems that the present meson-exchange approach dose
not support the states $D^{(*)}D^{(*)}[I(J^P)=0(2^+),1(0^+),1(1^+),1(2^+)]$
and $D^{(*)}\bar{B}^{(*)}[I(J^P)=1(1^+)]$ to be molecules.
Despite the \textit{reasonable} cutoff, we can not draw a definite
conclusion about the state $\bar{B}^{(*)}\bar{B}^{(*)}[I(J^P)=0(2^+)]$
due to the strong dependence on the cutoff. Further detailed study with
other approaches will be very helpful to settle this issue.

In the $S=1$ sector, the states $(\bar{B}^{(*)}\bar{B}^{(*)})_s[J^P=1^+,2^+]$ might be good molecule candidate. The states
$(D^{(*)}D^{(*)})_s[J^P=1^+]$, $(\bar{B}^{(*)}\bar{B}^{(*)})_s[J^P=0^+]$ and $(D^{(*)}\bar{B}^{(*)})_s[J^P=2^+]$ also seems to be molecule candidates, but  not good ones because the results depend a little sensitively on the cutoff. With the same reason as for the $S=0$ case, we can not draw definite conclusions on the states $(D^{(*)}\bar{B}^{(*)})_s[J^P=0^+,1^+]$. However, the present meson-exchange approach does not seem to support states $(D^{(*)}D^{(*)})_s[J^P=0^+,2^+]$ to be molecules.

For the $S=2$ systems, our results suggest that the states
$(\bar{B}^{(*)}\bar{B}^{(*)})_{ss}[J^P=1^+,2^+]$ might be good molecule candidates whereas the states $(D^{(*)}D^{(*)})_{ss}[J^P=0^+,1^+,2^+]$ might not be molecules. Although the results depend a little sensitively on the cutoff, the
states $(\bar{B}^{(*)}\bar{B}^{(*)})_{ss}[J^p=0^+]$ and $(D^{(*)}\bar{B}^{(*)})_{ss}[J^P=0^+,1^+,2^+]$ might also be molecules. We summarize our conclusions in Table~\ref{summation}.

In Reference~\cite{Molina:2010tx}, the authors also
studied the doubly charmed systems within the hidden gauge formalism
in a coupled-channel unitary approach. For the $D^*D^*$ systems with
$C=2, S=0$ and $I=0$, they only obtained a bound state with quantum
number $I(J^P)=0(1^+)$, which is similar to our result. However, the
pole appeared at 3969 MeV, which is 100 MeV larger than our result,
3870 MeV. This is because we consider the coupled channel effect of
$DD^*$ to $D^*D^*$. Actually, what we obtain is a $DD^*$ bound
state, but not a $D^*D^*$ bound state. For the systems with $C=2,
S=0$ and $I=1$, both their results and ours indicate that there are
no molecule candidates. For the systems with $C=2, S=1$ and
$I={1\over 2}$, they only obtained a bound state with quantum
numbers $I(J^P)={1\over 2}(1^+)$. Our results indicate that only the
state with $I(J^P)={1\over 2}(1^+)$ might be a molecule candidate,
but not a ideal one because the result depends a bit sensitively on
the cutoff parameter. For the systems with $C=2, S=2$ and $I=0$,
they obtained no bound states, neither do we, see
Table~\ref{summation}. Our results and those in
Reference~\cite{Molina:2010tx}are consistent with each other,
although these two theoretical frameworks were quite different.

It is very interesting to search for the predicted exotic hadronic
molecular states experimentally. These molecular candidates cannot
directly fall apart into the corresponding components due to the
absence of the phase space. For these molecular states with double
charm, they cannot decay into a double charm baryon plus a light
baryon. The masses of the lightest doubly-charmed baryon and light
baryon are 3518 MeV and 938 MeV, respectively, corresponding to
$\Xi_{cc}^+$ and proton as listed in PDG
\cite{0954-3899-37-7A-075021}. The mass of the molecular state is
around 3850 MeV and much smaller than the sum of the masses of a
doubly-charmed baryon and a light baryon. Therefore such a decay
is kinematically forbidden.

However, the heavy vector meson within the exotic molecular state
may decay. The $D^*$ mainly decays to $D\pi$ via strong
interaction. It also decays into $D\gamma$. Similarly, $D_s^*$,
$B^*$ and $B_s^*$ dominantly decay to $D_s\gamma$, $B\gamma$ and
$B_s\gamma$ via electromagnetic interaction, respectively. For
example, the main decay modes of the exotic double-charm molecular
state with one $D^*$ meson are $DD\gamma$ and $DD\pi$. If the
molecular candidates contain two heavy pseudoscalar mesons only,
they are stable once produced. The $D^*$ meson may also
decay via weak interaction. For these exotic molecules with double
bottom or both one charm and one bottom, their decay behavior is
similar to that of the molecular state with double charm.

The above typical decay modes provide important information to
further experimental search. Although very difficult, it is still
possible to produce such heavy systems with double bottom or both
one charm and one bottom at LHC.

\begin{table*}[htp]
\centering
\caption{The summary of our conclusions. ``$***$" means the corresponding state dose not exist due to symmetry. ``$\surd$" (``$\times$") means the corresponding state might ( might not) be a molecule. ``$\hright$" denotes that the corresponding state might be a molecule candidate, but not a good one because the results depend a little sensitive on the cutoff. However,``$?$" means we can not draw a definite conclusion with the present meson-exchange approach}

\label{summation}
\begin{tabular*}{18cm}{@{\extracolsep{\fill}}cccccccccc}
\toprule[1.0pt]\addlinespace[3pt]
     ~     &\multicolumn{3}{c}{Double Charm}       &
\multicolumn{3}{c}{Double Bottom} & \multicolumn{3}{c}{Charm \& Bottom} \\
Strangeness&      ~       &  $I(J^P)$    & Status  &   ~
     ~     &  $I(J^P)$    &  Status      &     ~   & $I(J^P)$ &  Status \\
\hline
\multirow{6}*{$S=0$}&\multirow{6}*{$D^{(*)}D^{(*)}$}&$0(0^+)$&$***$&
\multirow{6}*{$\bar{B}^{(*)}\bar{B}^{(*)}$}&$0(0^+)$    & $***$     &
\multirow{6}*{$D^{(*)}\bar{B}^{(*)}$}& $0(0^+)$         & $\hright$    \\
       ~   &     ~       & $0(1^+)$      &$\surd $      &  ~   &$0(1^+)$&
$\surd$    &     ~       & $0(1^+)$      &$\surd $ \\
      ~    &     ~       & $0(2^+)$      &$\times$ &    ~   & $0(2^+)$ &
?          &     ~       & $0(2^+)$      &$\surd$  \\
    ~      &     ~       & $1(0^+)$      &$\times$ &    ~   & $1(0^+)$ &
$\hright$  &     ~       & $1(0^+)$      &$\hright$ \\
     ~     &     ~       & $1(1^+)$      &$\times$ &    ~   &  $1(1^+)$ &
$\surd$    &     ~       & $1(1^+)$      &$\times$  \\
    ~      &     ~       & $1(2^+)$      &$\times$ &    ~   & $1(2^+)$&
$\hright$  &     ~       & $1(2^+)$      &$\hright$ \\
 \hline
\multirow{3}*{$S=1$}     &\multirow{3}*{$(D^{(*)}D^{(*)})_s$}
&$\frac{1}{2}(0^+)$&$\times$&\multirow{3}*{$(\bar{B}^{(*)}\bar{B}^{(*)})_s$}&
$\frac{1}{2}(0^+)$ &$\hright$&\multirow{3}*{$(D^{(*)}\bar{B}^{(*)})_s$}      &$\frac{1}{2}(0^+)$      & ?       \\
~          &             &$\frac{1}{2}(1^+)$ &$\hright$&    ~   &
$\frac{1}{2}(1^+)$       &$\surd$ &  ~       &$\frac{1}{2}(1^+)$& ? \\
   ~      &       ~      &$\frac{1}{2}(2^+)$ &$\times$&    ~   &
$\frac{1}{2}(2^+)$       &$\surd$          &  ~   &$\frac{1}{2}(2^+)$&
$\hright$      \\
\hline
\multirow{3}*{$S=-2$}    &\multirow{3}*{$(D^{(*)}D^{(*)})_{ss}$}&
$0(0^+)$  & $\times$     &\multirow{3}*{$(\bar{B}^{(*)}\bar{B}^{(*)})_{ss}$}&
$0(0^+)$  &$\hright$     &\multirow{3}*{$(D^{(*)}\bar{B}^{(*)})_{ss}$}&
$0(0^+)$  &$\hright$��\\
    ~     &    ~         & $0(1^+)$      &$\times$  &    ~  & $0(1^+)$&
$\surd$   &    ~         & $0(1^+)$      &$\hright$ \\
    ~     &    ~         & $0(2^+)$      &$\times$  &   ~   & $0(2^+)$&
$\surd$   &    ~         & $0(2^+)$      &$\hright$  \\
\bottomrule[1.0pt]
\end{tabular*}
\end{table*}

\section{ACKNOWLEDGEMENT}
This project was supported by the National Natural Science
Foundation of China under Grants 11075004, 11021092, 11261130311,
11222547, 11175073, 11035006, Ministry of Science and Technology
of China (2009CB825200), the Ministry of Education of China
(FANEDD under Grant No. 200924, DPFIHE under Grant No.
20090211120029, NCET, the Fundamental Research Funds for the
Central Universities) and the Fok Ying-Tong Education Foundation
(No. 131006). This work is also supported in part by the DFG and
the NSFC through funds provided to the sinogermen CRC 110
``Symmetries and the Emergence of Structure in QCD".


\section{APPENDIX}\label{Appendix}

The functions $H_i$ etc are defined as,
\begin{eqnarray}
H_0(\Lambda,q_0,m,r)&=&\frac{u}{4\pi}\left[Y(u
r)-\frac{\chi}{u}Y(\chi r)
-\f{r\beta^2}{2u}Y(\chi r)\right], \nonumber\\
H_1(\Lambda,q_0,m,r)&=&\frac{u^3}{4\pi}\left[Y(u
r)-\f{\chi}{u}Y(\chi r)
-\f{r\chi^2\beta^2}{2u^3}Y(\chi r)\right], \nonumber\\
H_3(\Lambda,q_0,m,r)&=&\frac{u^3}{12\pi}\left[Z(u r)-\f{\chi^3}{u^3}Z(\chi r) -\f{\chi\beta^2}{2u^3}Z_2(\chi r)\right],\nonumber \\
M_1(\Lambda,q_0,m,r)&=&-\frac{u^3}{4\pi}\left\{\f{1}{\theta
r}\left[\cos(\theta r)-e^{-\chi r}\right]
+\frac{\chi\beta^2}{2\theta^3}e^{-\chi r}\right\},\nonumber\\
M_3(\Lambda,q_0,m,r)&=&-\frac{u^3}{12\pi}\left\{\left[\cos{(\theta r)}-\f{3\sin{(\theta r)}}{\theta r} -\f{3\cos{(\theta r)}}{\theta^2r^2}\right]\right.\nonumber\\
&~&\left.\times\f{1}{\theta r}+\f{\chi^3}{\theta^3}Z(\chi r)
+\frac{\chi\beta^2}{2\theta^3}Z_2(\chi r)\right\},
\end{eqnarray}
where,
\begin{eqnarray*}
 \beta^2=\L^2-m^2,&\quad& u^2=m^2-q_0^2, \nonumber\\
\theta^2=-(m^2-q_0^2),&\quad&\chi^2=\L^2-q_0^2,
\end{eqnarray*}
and
\begin{eqnarray*}
 Y(x)=\f{e^{-x}}{x},&\quad& Z(x)=\left(1+\f{3}{x}+\f{3}{x^2}\right)Y(x),\nonumber\\
 Z_1(x)=\left(\f{1}{x}+\f{1}{x^2}\right)Y(x),&\quad& Z_2(x)=(1+x)Y(x).
\end{eqnarray*}
Fourier transformation formulas read:
\begin{eqnarray}
\frac{1}{u^2+\bm{q}^2}&\rightarrow& H_0(\Lambda,q_0,m,r),\nonumber \\
\frac{\bm{q}^2}{u^2+\bm{q}^2}&\rightarrow&-H_1(\Lambda,q_0,m,r), \nonumber\\
\frac{q_iq_j}{u^2+\bm{q}^2}&\rightarrow&-H_3(\Lambda,q_0,m,r)
k_{ij}-\frac{1}{3}H_1(\Lambda,q_0,m,r)\delta_{ij}, \nonumber\\
\label{FTformula}
\end{eqnarray}
where, $k_{ij}=3\f{r_ir_j}{r^2}-\delta_{ij}.$

We summarize the isospin-dependent coefficients in
Tables~\ref{coefficient:S0},\ref{coefficient:S1},\ref{coefficient:S2}
and the time component of the transferred momentum used in our
calculation in Table~\ref{q0}.

\begin{table*}
\renewcommand{\arraystretch}{1.45}
 \caption{The isospin-dependent coefficients for the $D^{(*)}D^{(*)}$ and $D^{(*)}\bar{B}^{(*)}$ systems with strangeness $S=0$. The coefficients of the $\bar{B}^{(*)}\bar{B}^{(*)}$ systems, which are not shown, are similar to those of the $D^{(*)}D^{(*)}$.}\label{coefficient:S0}
\begin{tabular*}{18cm}{@{\extracolsep{\fill}}ccccc}
\hline
    $I=0$ &   $DD$  & $[DD^*]_{-}$     &  $[DD^*]_{+}$      &  $D^*D^*$ \\
\hline
    $DD$  &$-\frac{3}{2}\rho^a+\frac{1}{2}\omega^a+\sigma^a$ &           &
         &$-\frac{3}{2}\pi^c+\frac{1}{6}\eta^c-\frac{3}{2}\rho^c+\frac{1}{2}\omega^c$ \\
$[DD^*]_{-}$&
&$-\frac{3}{2}\rho^d+\frac{1}{2}\omega^d+\sigma^d$&     &
$\frac{1}{\sqrt{2}}(-\frac{3}{2}\pi^f+\frac{1}{6}\eta^f-\frac{3}{2}\rho^f+\frac{1}{2}\omega^f)$\\
&         &
$-(-\frac{3}{2}\pi^e+\frac{1}{6}\eta^e-\frac{3}{2}\rho^e+\frac{1}{2}\omega^e)$
&         &
$-\frac{1}{\sqrt{2}}(-\frac{3}{2}\pi^g+\frac{1}{6}\eta^g-\frac{3}{2}\rho^g+\frac{1}{2}\omega^g)$\\
$[DD^*]_{+}$&       &           & $-\frac{3}{2}\rho^d+\frac{1}{2}\omega^d+\sigma^d$& $\frac{1}{\sqrt{2}}(-\frac{3}{2}\pi^f+\frac{1}{6}\eta^f-\frac{3}{2}\rho^f+\frac{1}{2}\omega^f)$\\
            &       &           &
$+(-\frac{3}{2}\pi^e+\frac{1}{6}\eta^e-\frac{3}{2}\rho^e+\frac{1}{2}\omega^e)$
&$+\frac{1}{\sqrt{2}}(-\frac{3}{2}\pi^g+\frac{1}{6}\eta^g-\frac{3}{2}\rho^g+\frac{1}{2}\omega^g)$\\
$D^*D^*$    &       &           &               &
$-\frac{3}{2}\pi^h+\frac{1}{6}\eta^h-\frac{3}{2}\rho^h+\frac{1}{2}\omega^h+\sigma^h$\\
\hline
    $I=1$ &   $DD$  & $[DD^*]_{-}$     &  $[DD^*]_{+}$      &  $D^*D^*$ \\
\hline
    $DD$  &$\frac{1}{2}\rho^a+\frac{1}{2}\omega^a+\sigma^a$ &           &
         &$\frac{1}{2}\pi^c+\frac{1}{6}\eta^c+\frac{1}{2}\rho^c+\frac{1}{2}\omega^c$ \\
$[DD^*]_{-}$&
&$\frac{1}{2}\rho^d+\frac{1}{2}\omega^d+\sigma^d$&     &
$\frac{1}{\sqrt{2}}(\frac{1}{2}\pi^f+\frac{1}{6}\eta^f+\frac{1}{2}\rho^f+\frac{1}{2}\omega^f)$\\
&         &
$-(\frac{1}{2}\pi^e+\frac{1}{6}\eta^e+\frac{1}{2}\rho^e+\frac{1}{2}\omega^e)$
&         &
$-\frac{1}{\sqrt{2}}(\frac{1}{2}\pi^g+\frac{1}{6}\eta^g+\frac{1}{2}\rho^g+\frac{1}{2}\omega^g)$\\
$[DD^*]_{+}$&       &           & $\frac{1}{2}\rho^d+\frac{1}{2}\omega^d+\sigma^d$& $\frac{1}{\sqrt{2}}(\frac{1}{2}\pi^f+\frac{1}{6}\eta^f+\frac{1}{2}\rho^f+\frac{1}{2}\omega^f)$\\
            &       &           &
$+(\frac{1}{2}\pi^e+\frac{1}{6}\eta^e+\frac{1}{2}\rho^e+\frac{1}{2}\omega^e)$
&$+\frac{1}{\sqrt{2}}(\frac{1}{2}\pi^g+\frac{1}{6}\eta^g+\frac{1}{2}\rho^g+\frac{1}{2}\omega^g)$\\
$D^*D^*$    &       &           &               &
$\frac{1}{2}\pi^h+\frac{1}{6}\eta^h+\frac{1}{2}\rho^h+\frac{1}{2}\omega^h+\sigma^h$\\
\hline
    $I=0$ & $D\bar{B}$ & $D\bar{B}^*$&$D^*\bar{B}$&$D^*\bar{B}^*$
\\ \hline
$D\bar{B}$&$-\frac{3}{2}\rho^a+\frac{1}{2}\omega^a+\sigma^a$ &
&
          &$-\frac{3}{2}\pi^c+\frac{1}{6}\eta^c-\frac{3}{2}\rho^c+\frac{1}{2}\omega^c$ \\
$D\bar{B}^*$&
&$-\frac{3}{2}\rho^d+\frac{1}{2}\omega^d+\sigma^d$
&$-\frac{3}{2}\pi^e+\frac{1}{6}\eta^e-\frac{3}{2}\rho^e+\frac{1}{2}\omega^e$&
$-\frac{3}{2}\pi^f+\frac{1}{6}\eta^f-\frac{3}{2}\rho^f+\frac{1}{2}\omega^f$\\
$D^*\bar{B}$ &        &          &
$-\frac{3}{2}\rho^d+\frac{1}{2}\omega^d+\sigma^d$ &
$-\frac{3}{2}\pi^g+\frac{1}{6}\eta^g-\frac{3}{2}\rho^g+\frac{1}{2}\omega^g$\\
$D^*\bar{B}^*$&       &           &          &
$-\frac{3}{2}\pi^h+\frac{1}{6}\eta^h-\frac{3}{2}\rho^h+\frac{1}{2}\omega^h+
\sigma^h$\\
\hline
   $I=1$ & $D\bar{B}$ & $D\bar{B}^*$&$D^*\bar{B}$&$D^*\bar{B}^*$
\\ \hline
$D\bar{B}$&$\frac{1}{2}\rho^a+\frac{1}{2}\omega^a+\sigma^a$ &
&
          &$\frac{1}{2}\pi^c+\frac{1}{6}\eta^c+\frac{1}{2}\rho^c+\frac{1}{2}\omega^c$ \\
$D\bar{B}^*$&
&$\frac{1}{2}\rho^d+\frac{1}{2}\omega^d+\sigma^d$
&$\frac{1}{2}\pi^e+\frac{1}{6}\eta^e+\frac{1}{2}\rho^e+\frac{1}{2}\omega^e$&
$\frac{1}{2}\pi^f+\frac{1}{6}\eta^f+\frac{1}{2}\rho^f+\frac{1}{2}\omega^f$\\
$D^*\bar{B}$ &        &          &
$\frac{1}{2}\rho^d+\frac{1}{2}\omega^d+\sigma^d$ &
$\frac{1}{2}\pi^g+\frac{1}{6}\eta^g+\frac{1}{2}\rho^g+\frac{1}{2}\omega^g$\\
$D^*\bar{B}^*$&       &           &          &
$\frac{1}{2}\pi^h+\frac{1}{6}\eta^h+\frac{1}{2}\rho^h+\frac{1}{2}\omega^h+
\sigma^h$\\
\hline
\end{tabular*}
\end{table*}
\begin{table*}
\renewcommand{\arraystretch}{1.45}
 \caption{The isospin-dependent coefficients for the $(D^{(*)}D^{(*)})_s$ and $(D^{(*)}\bar{B}^{(*)})_s$ systems with strangeness $S=1$. The coefficients of the $(\bar{B}^{(*)}\bar{B}^{(*)})_s$ systems, which are not shown, are similar to those of the $(D^{(*)}D^{(*)})_s$.}\label{coefficient:S1}
\begin{tabular*}{18cm}{@{\extracolsep{\fill}}ccccccccc}
\hline
          &\multicolumn{2}{c}{$DD_s$}&\multicolumn{2}{c}{$DD_s^*$}&
          \multicolumn{2}{c}{$D^*D_s$}&\multicolumn{2}{c}{$D^*D_s^*$}\\
\hline
$DD_s$   &\multicolumn{2}{c}{$\sigma^a+K^{*a}$}&\multicolumn{2}{c}{0}&\multicolumn{2}{c}{0}     &\multicolumn{2}{c}{$-\frac{1}{3}\eta^c+K^c+K^{*c}$}\\
$DD_s^*$ &      &
&\multicolumn{2}{c}{$\sigma^d+K^e+K^{*e}$}&
\multicolumn{2}{c}{$-\frac{1}{3}\eta^e+K^{*d}$}&\multicolumn{2}{c}{$-\frac{1}{3}\eta^f+K^f+K^{*f}$}\\
$D^*D_s$ &   &    &     &      & \multicolumn{2}{c}{$\sigma^d+K^e+K^{*e}$}& \multicolumn{2}{c}{$-\frac{1}{3}\eta^g+K^g+K^{*g}$}\\
$D^*D_s^*$    &   &   &     &      &       &        &
\multicolumn{2}{c}{$-\frac{1}{3}\eta^h+\sigma^h+K^h+K^{*h}$}\\
\hline
           &$D\bar{B}_s$ &$D_s\bar{B}$ & $D\bar{B}_s^*$ & $D_s^*\bar{B}$&
$D^*\bar{B}_s$ & $D_s\bar{B}^*$    &  $D^*\bar{B}_s^*$ & $D_s^*\bar{B}^*$ \\
\hline $D\bar{B}_s$&$\sigma^a$ &$K^{*a}$&  0   &   0  &    0     &
0     &
$-\frac{1}{3}\eta^c$    &$K^c+K^{*c}$ \\
$D_s\bar{B}$ &           &$\sigma^a$&  0   &   0  &    0     &   0
&
$K^c+K^{*c}$&$-\frac{1}{3}\eta^c$\\
$D\bar{B}_s^*$&         &
&$\sigma^d$&$K^e+K^{*e}$&$-\frac{1}{3}\eta^{*e}$&$K^{*d}$&
$-\frac{1}{3}\eta^f$    &$K^f+K^{*f}$\\
$D_s^*\bar{B}$&         &        &      &$\sigma^d$& $K^{*d}$
&$-\frac{1}{3}\eta^e$    &$K^g+K^{*g}$&$-\frac{1}{3}\eta^g$\\
$D^*\bar{B}_s$&         &        &      &          &$\sigma^d$
&
$K^e+K^{*e}$  &$-\frac{1}{3}\eta^g$&$K^g+K^{*g}$  \\
$D_s\bar{B}^*$&          &        &      &          &
&
 $\sigma^d$   &$K^f+K^{*f}$&$-\frac{1}{3}\eta^f$\\
$D^*\bar{B}_s^*$&        &      &          &        &        &
&
$-\frac{1}{3}\eta^h+\sigma^h$&$K^h+K^{*h}$\\
$D_s^*\bar{B}^*$&        &        &      &          &
&
&                        &$-\frac{1}{3}\eta^h+\sigma^h$\\
\hline
\end{tabular*}
\end{table*}
\begin{table*}
\renewcommand{\arraystretch}{1.5}
 \caption{The isospin-dependent coefficients for the $(D^{(*)}D^{(*)})_{ss}$ and $(D^{(*)}\bar{B}^{(*)})_{ss}$ systems with strangeness $S=1$. The coefficients of the $(\bar{B}^{(*)}\bar{B}^{(*)})_{ss}$ systems, which are not shown, are similar to those of the $(D^{(*)}D^{(*)})_{ss}$.}\label{coefficient:S2}
\begin{tabular*}{18cm}{@{\extracolsep{\fill}}ccccc}
\hline
    $I=1$ &   $D_sD_s$  & $[D_sD_s^*]_{-}$&$[D_sD_s^*]_{+}$ &$D_s^*D_s^*$ \\
\hline
$D_sD_s$  &$\phi^a+\sigma^a$ &     0      &     0   &$\frac{2}{3}\eta^c+\phi^c$ \\
$[D_sD_s^*]_{-}$&
&$\phi^d+\sigma^d-(\frac{2}{3}\eta^e+\phi^e)$&$\times$&
$\frac{1}{\sqrt{2}}(\frac{2}{3}\eta^f+\phi^f)-\frac{1}{\sqrt{2}}(\frac{2}{3}\eta^g+\phi^g)$\\
$[D_sD_s^*]_{+}$&       &           & $\phi^d+\sigma^d+(\frac{2}{3}\eta^e+\phi^e)$& $\frac{1}{\sqrt{2}}(\frac{2}{3}\eta^f+\phi^f)+\frac{1}{\sqrt{2}}(\frac{2}{3}\eta^g+\phi^g)$\\
$D_s^*D_s^*$    &       &           &               &
$\frac{2}{3}\eta^h+\phi^h+\sigma^h$\\
\hline
           &$D_s\bar{B}_s$ & $D_s\bar{B}_s^*$ & $D_s^*\bar{B}_s$&  $D_s^*\bar{B}_s^*$ \\
\hline
$D_s\bar{B}_s$&$\phi^a+\sigma^a$ &     0      &     0      &$\frac{2}{3}\eta^c+\phi^c$\\
$D_s\bar{B}_s^*$&
&$\phi^d+\sigma^d$&$\frac{2}{3}\eta^e+\phi^e$     &
$\frac{2}{3}\eta^f+\phi^f$\\
$D_s^*\bar{B}_s$&       &           & $\phi^d+\sigma^d$& $\frac{2}{3}\eta^g+\phi^g$\\
$D_s^*\bar{B}_s^*$  &           &           &      &
$\frac{2}{3}\eta^h+\phi^h+\sigma^h$\\
\hline

\end{tabular*}
\end{table*}
\begin{table*}[htp]
\renewcommand{\arraystretch}{1.5} \caption{The time component of the transferred momentum, $q_0$, used in our calculation. The other values which are not given are zero. }\label{q0}
\begin{tabular*}{18cm}{@{\extracolsep{\fill}}cccccccc}
\toprule[0.8pt] \toprule[0.8pt] \addlinespace[2pt]
 \multicolumn{2}{c}{$DD$} & \multicolumn{2}{c}{$\bar{B}\bar{B}$}&\multicolumn{4}{c}{$D\bar{B}$} \\
Process   &  $q_0$     & Process     &   $q_0$   &  Process    &  $q_0$  &  Process   &  $q_0$    \\
\specialrule{0.6pt}{1pt}{3pt} $DD_s\to D_sD$  &$m_{D_s}-m_D$
&$\bar{B}\bar{B}_s\to\bar{B}_s\bar{B}$ &$m_{B_s}-m_B$
&$D_s\bar{B}\to D\bar{B}_s$
&$\frac{(m_{B_s}^2+m_{D_s}^2)-(m_B^2+m_D^2)}{2(m_D+m_{B_s})}$
&$D\bar{B}_s\to D_s\bar{B}$
&$\frac{(m_{B_s}^2+m_{D_s}^2)-(m_B^2+m_D^2)}{2(m_D+m_{B_s})}$\\
$D_sD\to D_s^*D^*$
&$\frac{(m_{D_s^*}^2+m_D^2)-(m_{D_s}^2+m_{D^*}^2)}{2(m_{D_s^*}+m_{D^*})}$
&$\bar{B}_s\bar{B}\to \bar{B}_s^*\bar{B}^*$
&$\frac{(m_{B_s^*}^2+m_B^2)-(m_{B_s}^2+m_{B^*}^2)}{2(m_{B_s^*}+m_{B^*})}$
&$D_s\bar{B}\to D_s^*\bar{B}^*$
&$\frac{(m_{D_s^*}^2+m_B^2)-(m_{B^*}^2+m_{D_s}^2)}{2(m_{D_s^*}+m_{B^*})}$
&$D_s\bar{B}\to D^*\bar{B}_s^*$
&$\frac{(m_{B_s^*}^2+m_{D_s}^2)-(m_{D^*}^2+m_B^2)}{2(m_{D^*}+m_{B_s^*})}$\\
$D_sD\to D^*D_s^*$
&$\frac{(m_{D_s^*}^2+m_{D_s}^2)-(m_{D}^2+m_{D^*}^2)}{2(m_{D^*}+m_{D_s^*})}$
&$\bar{B}_s\bar{B}\to\bar{B}^*\bar{B}_s^*$ &$
\frac{(m_{B_s^*}^2+m_{B_s}^2)-(m_{B^*}^2+m_B^2)}{2(m_{B^*}+m_{B_s^*})}$
&$D\bar{B}_s\to D_s^*\bar{B}^*$
&$\frac{(m_{D_s^*}^2+m_{B_s}^2)-(m_{B^*}^2+m_D^2)}{2(m_{D_s^*}+m_{B^*})}$
&$D\bar{B}_s\to D^*\bar{B}_s^*$
&$\frac{(m_{D^*}^2+m_{B_s}^2)-(m_{B_s^*}^2+m_D^2)}{2(m_{D^*}+m_{B_s^*})}$\\
$D_sD^*\to DD_s^*$
&$\frac{(m_{D_s^*}^2+m_{D_s}^2)-(m_D^2+m_{D^*}^2)}{2(m_D+m_{D_s^*})}$
&$\bar{B}_s\bar{B}^*\to\bar{B}\bar{B}_s^*$
&$\frac{(m_{B_s^*}^2+m_{B_s}^2)-(m_B^2+m_{B^*}^2)}{2(m_B+m_{B_s^*})}$
&$D_s\bar{B}^*\to D\bar{B}_s^*$
&$\frac{(m_{D_s}^2+m_{B_s^*}^2)-(m_D^2+m_{B^*}^2)}{2(m_{D_s}+m_{B^*})}$
&$D_s^*\bar{B}\to D^*\bar{B}_s$
&$\frac{(m_{D_s^*}^2+m_{B_s}^2)-(m_B^2+m_{D^*}^2)}{2(m_{D^*}+m_{B_s})}$\\
$DD^*\to D^*D$ &$m_{D^*}-m_D$
&$\bar{B}\bar{B}^*\to\bar{B}^*\bar{B}$ &$m_{B^*}-m_B$
&$D\bar{B}^*\to D^*\bar{B}$
&$\frac{(m_{D^*}^2-m_{B^*}^2)-(m_D^2+m_B^2)}{2(m_{D^*}+m_B)}$
&$D^*\bar{B}\to D\bar{B}^*$
&$\frac{(m_{D^*}^2-m_{B^*}^2)-(m_D^2+m_B^2)}{2(m_{D^*}+m_B)}$\\
$D_sD^*\to D_s^*D$
&$\frac{(m_{D_s^*}^2+m_{D^*}^2)-(m_{D_s}^2+m_D^2)}{2(m_{D_s}+m_{D^*})}$
&$\bar{B}_s\bar{B}^*\to\bar{B}_s^*\bar{B}$
&$\frac{(m_{B_s^*}^2+m_{B^*}^2)-(m_B^2+m_{B_s}^2)}{2(m_{B_s}+m_{B^*})}$
&$D_s\bar{B}^*\to D_s^*\bar{B}$
&$\frac{(m_{D_s^*}^2+m_{B^*}^2)-(m_{D_s}^2+m_B^2)}{2(m_{D_s^*}+m_B)}$
&$D_s^*\bar{B}\to D_s\bar{B}^*$
&$\frac{(m_{B^*}^2+m_{D_s^*}^2)-(m_B^2+m_D^2)}{2(m_{D_s^*}+m_B)}$\\
$D_sD^*\to D^*D_s$ &$m_{D^*}-m_{D_s}$ &$\bar{B}_s\bar{B}^*\to
\bar{B}^*\bar{B}_s$ &$m_{B^*}-m_{B_s}$ &$D_s\bar{B}^*\to
D^*\bar{B}_s$
&$\frac{(m_{B^*}^2+m_{D^*}^2)-(m_{B_s}^2+m_{D_s}^2)}{2(m_{D^*}+m_{B_s})}$
&$D_s^*\bar{B}\to D\bar{B}_s^*$
&$\frac{(m_{D_s^*}^2+m_{B_s^*}^2)-(m_D^2+m_B^2)}{2(m_{D_s^*}+m_B)}$\\
$DD_s^*\to D_s^*D$ &$m_{D_s^*}-m_D$ &$\bar{B}\bar{B}_s^*\to
\bar{B}_s^*\bar{B}$ &$m_{B_s^*}-m_B$ &$D\bar{B}_s^*\to
D_s^*\bar{B}$
&$\frac{(m_{B_s^*}^2+m_{D_s^*}^2)-(m_D^2+m_B^2)}{2(m_{D_s^*}+m_B)}$
&$D^*\bar{B}_s\to D_s\bar{B}^*$
&$\frac{(m_{D^*}^2+m_{B^*}^2)-(m_{B_s}^2+m_{D_s}^2)}{2(m_{D_s}+m_{B^*})}$\\
$DD_s^*\to D^*D_s$
&$\frac{(m_{D_s^*}^2+m_{D^*}^2)-(m_{D_s}^2-m_D^2)}{2(m_{D_s}+m_{D^*})}$
&$\bar{B}\bar{B}_s^*\to\bar{B}^*\bar{B}_s$
&$\frac{(m_{B_s^*}^2+m_{B^*}^2)-(m_B^2+m_{B_s}^2)}{2(m_{B^*}+m_{B_s})}$
&$D\bar{B}_s^*\to D^*\bar{B}_s$
&$\frac{(m_{B_s^*}^2+m_{D^*}^2)-(m_D^2+m_{B_s}^2)}{2(m_D+m_{B_s^*})}$
&$D^*\bar{B}_s\to D\bar{B}_s^*$
&$\frac{(m_{D^*}^2+m_{B_s^*}^2)-(m_D^2+m_{B_s}^2)}{2(m_D+m_{B_s^*})}$\\
$DD^*\to D^*D^*$ &$\frac{m_{D^*}^2-m_D^2}{4m_{D^*}}$
&$\bar{B}\bar{B}^*\to \bar{B}^*\bar{B}^*$
&$\frac{m_{B^*}^2-m_B^2}{4m_{B^*}}$ &$D\bar{B}^*\to D^*\bar{B}^*$
&$\frac{m_{D^*}^2-m_D^2}{2(m_{D^*}+m_{B^*})}$ &$D^*\bar{B}\to
D^*\bar{B}^*$
&$\frac{m_{B^*}^2-m_B^2}{2(m_{D^*}+m_{B^*})}$\\
$D_sD^*\to D_s^*D^*$
&$\frac{(m_{D_s^*}^2-m_{D_s}^2)}{2(m_{D_s^*}+m_{D^*})}$
&$\bar{B}_s\bar{B}^*\to \bar{B}_s^*\bar{B}^*$
&$\frac{m_{B_s^*}^2-m_{B_s}^2}{2(m_{B_s^*}+m_{B^*})}$
&$D_s\bar{B}^*\to D_s^*\bar{B}^*$
&$\frac{m_{D_s^*}^2-m_{D_s}^2}{2(m_{D_s^*}+m_{B^*})}$
&$D_s^*\bar{B}\to D_s^*\bar{B}^*$
&$\frac{m_{B^*}^2-m_B^2}{2(m_{D_s^*}+m_{B^*})}$\\
$D_sD^*\to D^*D_s^*$
&$\frac{m_{D_s^*}^2+m_{D_s}^2-2m_{D^*}^2}{2(m_{D^*}+m_{D_s^*})}$
&$\bar{B}_s\bar{B}^*\to\bar{B}^*\bar{B}_s^*$
&$\frac{m_{B_s^*}^2+m_{B_s}^2-2m_{B^*}^2}{2(m_{B^*}+m_{B_s^*})}$
&$D_s\bar{B}^*\to D^*\bar{B}_s^*$
&$\frac{(m_{D^*}^2+m_{B^*}^2)-(m_{D_s}^2+m_{B_s^*}^2)}{2(m_{D^*}+m_{B_s^*})}$
&$D_s^*\bar{B}\to D^*\bar{B}_s^*$
&$\frac{(m_{D_s^*}^2+m_{B_s^*}^2)-(m_{D^*}^2+m_{B}^2)}{2(m_{D^*}+m_{B_s^*})}$\\
$DD_s^*\to D_s^*D^*$
&$\frac{2m_{D_s^*}^2-m_D^2-m_{D^*}^2}{2(m_{D_s^*}+m_{D^*})}$
&$\bar{B}\bar{B}_s^*\to\bar{B}_s^*\bar{B}^*$
&$\frac{2m_{B_s^*}^2-m_B^2-m_{B^*}^2}{2(m_{B_s^*}+m_{B^*})}$
&$D\bar{B}_s^*\to D_s^*\bar{B}^*$
&$\frac{(m_{D_s^*}^2+m_{B_s^*}^2)-(m_D^2+m_{B^*}^2)}{2(m_{D_s^*}+m_{B^*})}$
&$D^*\bar{B}_s\to D_s^*\bar{B}^*$
&$\frac{(m_{D_s^*}^2+m_{B_s}^2)-(m_{D^*}^2+m_{B^*}^2)}{2(m_{D_s^*}+m_{B^*})}$\\
$DD_s^*\to D^*D_s^*$
&$\frac{m_{D^*}^2-m_D^2}{2(m_{D^*}+m_{D_s^*})}$
&$\bar{B}\bar{B}_s^*\to\bar{B}^*\bar{B}_s^*$
&$\frac{m_{B^*}^2-m_B^2}{2(m_{B^*}+m_{B_s^*})}$ &$D\bar{B}_s^*\to
D^*\bar{B}_s^*$ &$\frac{m_{D^*}^2-m_D^2}{2(m_{D^*}+m_{B_s^*})}$
&$D^*\bar{B}_s\to D^*\bar{B}_s^*$
&$\frac{m_{B_s^*}^2-m_{B_s}^2}{2(m_{D^*}+m_{B_s^*})}$\\
$D_s^*D^*\to D^*D_s^*$ &$m_{D_s^*}-m_{D^*}$
&$\bar{B}_s^*\bar{B}^*\to\bar{B}^*\bar{B}_s^*$
&$m_{B_s^*}-m_{B^*}$ &$D_s^*\bar{B}^*\to D^*\bar{B}_s^*$
&$\frac{(m_{B_s^*}^2+m_{D_s^*}^2)-(m_{B^*}^2+m_{D^*}^2)}{2(m_{D^*}+m_{B_s^*})}$
&$D^*\bar{B}_s^*\to D_s^*\bar{B}^*$
&$\frac{(m_{B_s^*}^2+m_{D_s^*}^2)-(m_{B^*}^2+m_{D^*}^2)}{2(m_{D^*}+m_{B_s^*})}$\\
\bottomrule[0.8pt] \bottomrule[0.8pt]
\end{tabular*}
\end{table*}


\end{document}